\documentclass[fleqn,usenatbib]{mnras}

\usepackage{newtxtext,newtxmath}

\usepackage[T1]{fontenc}

\DeclareRobustCommand{\VAN}[3]{#2}
\let\VANthebibliography\thebibliography
\def\thebibliography{\DeclareRobustCommand{\VAN}[3]{##3}\VANthebibliography}

\usepackage{graphicx}	
\usepackage{amsmath}	
\usepackage{multirow}
\usepackage{gensymb}

\title[Black hole spin in PG 1535+547]{Constraining black hole spin in PG 1535+547 amidst complex multi-layered absorption}

\author[A. Madathil-Pottayil et al.]{
A. Madathil-Pottayil$^{1}$\thanks{E-mail: athulyamp9425@gmail.com}, 
D.J. Walton$^{1}$, 
Jiachen Jiang$^{2}$, T. Dauser$^{3}$, 
Andrew Fabian$^{4}$, 
D. Stern$^{5}$, \newauthor
Luigi C. Gallo$^{6}$, 
Mark T. Reynolds$^{7,8}$, 
Emanuele Nardini$^{9}$,
Javier A. Garcia$^{10}$\\
\\
$^{1}$Centre for Astrophysics Research, Department of Physics, Astronomy and Mathematics, University of Hertfordshire, College Lane, Hatfield AL10 9AB, UK\\
$^{2}$Department of Physics, University of Warwick, Gibbet Hill Road, Coventry CV4 7AL, UK\\
$^{3}$Dr. Karl-Remeis-Sternwarte and ECAP, Friedrich-Alexander-Universität Erlangen-Nürnberg, Sternwartstr. 7, 96049 Bamberg, Germany\\
$^{4}$Institute of Astronomy, University of Cambridge, Madingley Road, Cambridge CB3 0HA, UK\\
$^{5}$Jet Propulsion Laboratory, California Institute of Technology, 4800 Oak Grove Drive, Mail Stop 264-789, Pasadena, CA 91109, USA\\
$^{6}$Department of Astronomy and Physics, Saint Mary’s University, 923 Robie Street, Halifax, NS B3H 3C3, Canada\\
$^{7}$Department of Astronomy, Ohio State University, 140 W. 18th Avenue, Columbus, OH 43210, USA\\
$^{8}$Department of Astronomy, University of Michigan, 1085 S. University Ave., Ann Arbor, MI 48109, USA\\
$^{9}$INAF – Osservatorio Astrofisico di Arcetri, Largo Enrico Fermi 5, 50125 Firenze, Italy\\
$^{10}$NASA Goddard Space Flight Center, Greenbelt, MD 20771, USA\\}
\pubyear{\the\year{}}

\begin{document}
\label{firstpage}
\pagerange{\pageref{firstpage}--\pageref{lastpage}}
\maketitle

\begin{abstract}
We present a spectroscopic analysis of \textit{XMM-Newton} and \textit{NuSTAR} observations of the `complex’ NLS1 PG~1535+547 at redshift $z=0.038$. These observations span three epochs: 2002 and 2006 with \textit{XMM-Newton} alone, {covering the $0.3 - 10$ keV energy range,} and a coordinated \textit{XMM-Newton} and \textit{NuSTAR} observation in 2016, covering the $0.3-60$ keV energy range. The X-ray spectra across all epochs exhibit both neutral and ionized absorption, along with reflection features from the accretion disc, including a prominent Compton hump in the broadband data. Notably, the spectral shape varies across epochs. 
Our analysis suggests this variability is attributed to changes in both line-of-sight absorption and the intrinsic emission from PG~1535+547. The source is obscured by multiple layers of partially and/or fully covering neutral and ionized absorbers, with neutral column densities ranging from undetectable levels in the least obscured phase to $\sim0.3-5\times10^{23} ~ \rm cm^{-2} $ in the most obscured phase. A clear warm absorber is revealed during the least obscured phase. 
The continuum remains fairly consistent ($\Gamma \approx 2.2\pm0.1$) during the first two observations, followed by a substantial flux decrease (by a factor of $\sim7$ in the $2-10$ keV band) in 2016 compared to 2006. The 2016 data indicates the source is in a reflection-dominated state during this epoch, with a reflection fraction of  $R > 7$ and an X-ray source located at a height $\leq 1.72r_g$. Simultaneous fitting of the multi-epoch data suggests a rapidly rotating black hole with a spin parameter, $a > 0.99$. 
These findings imply that strong light-bending effects may account for the observed continuum flux reduction.
\end{abstract}

\begin{keywords}
black hole physics – galaxies: active – X-rays: individual: PG 1535+547
\end{keywords}



\section{Introduction}
\label{sec-1}
The spin of a black hole is one of the three fundamental properties essential for characterizing its nature, according to the no-hair theorem. 
Studying the relativistic broadening of the iron emission line is a well-established technique among the limited methods available to measure the spin. When Comptonized thermal photons reflect back and irradiate the accretion disc, they undergo scattering, absorption, and emission processes, thereby generating the reflection component observed in the X-ray spectrum of an AGN. This reflection spectrum exhibits prominent features such as the Fe K$\alpha$ line at $\sim6.4$ keV and the broad "Compton hump" between $\sim 10 - 30$ keV. The Fe K$\alpha$ line is strong owing to the high abundance and large fluorescence yield of iron \citep{1991MNRAS.249..352G}, while the Compton hump arises from the interplay between photoelectric absorption - which suppresses photons below $\sim 7 - 10$ keV due to the high photoelectric cross-section - and Compton scattering, which down-scatters photons at higher energies \citep{1988ApJ...335...57L,1988MNRAS.233..475G}.
This reflection component emerging from the inner regions of the accretion disc undergoes relativistic effects due to the strong gravitational potential of the supermassive black hole (SMBH). The extent of relativistic broadening on the reflection features correlates with the proximity of the disc to the innermost stable circular orbit (ISCO), which, in turn, correlates with the spin of the black hole \citep{2000PASP..112.1145F,2007ragt.meet..151M}. X-ray reflection spectroscopy therefore serves as a powerful tool to probe the spin of a black hole \citep{2006ApJ...652.1028B}.

The reflection scenario has been quite straightforward in explaining the broadened reflection features in unobscured  AGN \citep{2013MNRAS.428.2901W,2020MNRAS.497.2352M,2021A&A...654A..89P,2021ApJ...913...13X,2024MNRAS.534..608M}. However, extending the reflection scenario to moderately-obscured systems presents significant challenges. Despite the difficulties, it is essential to explore moderately-obscured systems to build large samples of AGN spin constraints in the future. This motivation stems from the scarcity of bright, bare AGN and the frequent observation of moderate levels of obscuration in AGN. Such obscuration is a common phenomenon observed in many type-1 Seyfert galaxies (Sy1s), where absorbing materials, such as accretion disc outflows \citep{2008A&A...483..437M} or gas clumps from within the Narrow and/or Broad Line Regions \citep{2013ApJ...766..104L}, align with our line-of-sight, partially or fully obstructing X-ray emission from the inner disc regions \citep{1997MNRAS.286..513R,1999ApJ...516..750C,2007A&A...463..131M,2010MNRAS.404.1369S,2013ApJ...766..104L,2023arXiv230210930G}. These absorbers often exhibit complex geometries and the absorption introduced by these structures can make it more difficult to accurately determine the profile of the iron emission from the disc sufficiently well to constrain the spin, since they can absorb away the red wing of the line profile.  
However, the broadband coverage provided by \textit{NuSTAR}  significantly helps in disentangling the spectroscopic signatures of absorption and reflection, as it offers sensitive coverage of the X-ray reflection spectrum extending to higher energies (up to 78 keV) and helps constrain the continuum with hard X-rays. 

In this work, we use the advanced relativistic reflection models along with the absorption models while allowing for a complex (and varying) line-of-sight absorption to understand the variability observed in the narrow-line Seyfert 1 (NLS1) PG 1535+547 and explore the potential of current relativistic reflection models to constrain black hole spin amidst complex line-of-sight absorption.

PG 1535+547 is a radio-quiet NLS1 galaxy with a narrow H$\beta$ line of width FWHM(H$\beta$) $= 1480$ km~s$^{-1}$. It was detected in the Palomar Green Bright Quasar Survey \citep{1983ApJ...269..352S} at a redshift of $z=0.038$ \citep{1991rc3..book.....D}. The mass of this source is estimated to be log($M_{\rm {BH}}/M_{\odot}$) $= 7.17$, derived from the FWHM(H$\beta$) line width (\citealt{2021ApJS..253...20H}; see also \citealt{2006ApJ...641..689V,2006ApJ...653..137Z}). PG\,1535+547 was initially recognized as a soft X-ray weak QSO ($\alpha_{\mathrm OX}$ $= -2.45$; \citealt{2000ApJ...528..637B}), where the soft X-ray luminosity is $\sim 10-30$ times weaker than expected based on its optical luminosity \citep{1997ApJ...477...93L}. The weakness in this source was attributed to heavy X-ray absorption with column densities $\geq 10^{22} \rm cm^{-2}$ \citep{2001ApJ...546..795G}. This source gained further attention due to its strong X-ray flux and spectral variability on timescales of months to years \citep{2005A&A...433..455S,2006MNRAS.368..479G,2008A&A...483..137B}. Previous analyses demonstrated this variability can be attributed to changes in the physical properties of warm absorbing gas along the line-of-sight, consistent with its classification as a mini-Broad Absorption Line (mini-BAL)\footnote{Mini-BAL is a subclass of AGN that exhibit strong, broad and blue-shifted absorption lines in their UV/X-ray spectra with absorption lines of intermediate width ($500~\rm km s^{-1} < FWHM < 2000~km s^{-1}$), falling between broad absorption line quasars (BALs)  and narrow absorption line quasars (NALs) \citep{1991ApJ...373...23W,2004ASPC..311..203H}} quasar \citep{2006RMxAA..42...23S}, where high-velocity winds are known to intercept the observer's line-of-sight.

PG 1535+547 has also exhibited a feature consistent with a broadened iron line. \cite{2005A&A...433..455S} proposed a relativistic reflection scenario to explain the broadening, while \cite{2008A&A...483..137B} suggested that it could instead be well described by a more elaborate absorption model, involving multiple zones of both warm and neutral absorbing material surrounding the AGN. 
In this study, we perform a broadband spectral analysis of the source using a recent, coordinated observation with \textit{XMM-Newton} and \textit{NuSTAR}, providing coverage over the $0.3 - 60$ keV bandpass. We also simultaneously perform a comparative analysis of the previous \textit{XMM-Newton} observations of the source, by homogeneously setting up the latest reflection models along with a complex line-of-sight absorption in all the observations.

The paper is structured as follows: In Section \ref{sec-2}, we describe the observations used in this study and outline the data reduction process. Section \ref{sec-3} presents our spectral analysis. We detail our results in Section \ref{sec-4}. Finally, we discuss our findings in Section \ref{sec-5} and summarize our conclusions in Section \ref{sec-6}.

\section{Observations and data reduction}
\label{sec-2}
In our study, we use the \textit{XMM-Newton} and \textit{NuSTAR} observations of the source taken during 2002, 2006 and 2016. A total of eight observations are apportioned into three epochs according to their observation times. Epochs 1 (Obs. 1) and 2 (Obs. 2, 3, and 4) comprise \textit{XMM-Newton} observations from 2002 and 2006, respectively. The coordinated \textit{XMM-Newton} and \textit{NuSTAR} observations from 2016, covering the broadband energy range from 0.3 to 60 keV, constitute epoch 3 (Obs. 5 - 8). A detailed log of all observations is provided in Table \ref{tab-1}. The reduction procedures for \textit{XMM-Newton} and \textit{NuSTAR} data are  outlined below.

\begin{table*}
\centering
\caption{Summary of observations of PG 1535+547}
\begin{tabular}{ccccccccccc}
\hline
Epoch & Obs. & Mission & Instrument &  Observation & Start Date & Exposure time$^a$  & Net src counts$^a$ & Scaled bkg counts$^a$\\
& & & & & & (ks) & (cts) & (cts) \\

\hline
1 & 1 & \textit{XMM-Newton} & pn \& MOS & 0150610301 & 2002-11-03 &  $14.6/34.4$ & $2067/1484$ & $274/80$ \\
\hline
\multirow{3}{*}{2} & 2 & \textit{XMM-Newton} & pn \& MOS & 0300310301 & 2006-01-16 &  $5.3/20.0$ & \multirow{3}{5em}{$14059/11230^*$} & \multirow{3}{3.5em}{$390/229^*$} \\
& 3 & \textit{XMM-Newton} & pn \& MOS & 0300310401 & 2006-01-22 &  $14.0/36.9$ &&\\
& 4 & \textit{XMM-Newton} & pn \& MOS & 0300310501 & 2006-01-24 &  $16.8/42.9$ &&\\
\hline

\multirow{4}{*}{3} & 5 & \textit{XMM-Newton} & pn \& MOS & 0790590101 & 2016-09-12 &  $30.2/72.2$ & \multirow{2}{4em}{$3168/2331^*$} & \multirow{2}{3.5em}{$294/203^*$}\\
& 6 & \textit{XMM-Newton} & pn \& MOS & 0790590201 & 2016-09-14 & $26.7/75.9$ && \\
& 7 & \textit{NuSTAR} & FPMA \& FPMB & 60201023002 & 2016-09-12 & $132.90$ & \multirow{2}{2em}{$3063^\dagger$} & \multirow{2}{1.5em}{$519^\dagger$}\\
& 8 & \textit{NuSTAR} & FPMA \& FPMB & 60201023004 & 2016-09-14 & $184.20$ && \\
\hline
\end{tabular}
\begin{flushleft}
$^a$ For the \textit{XMM-Newton} observations, values are listed in pn$/$MOS format. The MOS entries correspond to spectrum produced by merging MOS 1 and MOS 2 for each observation. \\
     $^{*}$ Net source counts and background counts scaled to the source-extraction region, in the $0.3 - 10$ keV band for the merged pn$/$MOS spectrum, obtained by merging all observations associated with the given epoch (see Section \ref{sec-2.1} for details).\\
     $^{\dagger}$ Net source and scaled background counts in the $3 - 60$ keV band for the merged \textit{NuSTAR} FPMA+FPMB spectrum (see Section \ref{sec-2.2} for details).
\end{flushleft}
\label{tab-1}
\end{table*}

\subsection{XMM-Newton}
\label{sec-2.1}
The \textit{XMM-Newton} observation data files (ODF) were obtained from the XMM-Newton Science Archive (XSA)\footnote{\url{https://www.cosmos.esa.int/web/xmm-newton/xsa}} and processed using its standard Science Analysis System (SAS). Both pn \citep{2001A&A...365L..27T} and MOS \citep{2001A&A...365L..18S} detectors were operating in ``full window mode". Clean event files for the pn and MOS data were generated using the \textsc{epproc} and \textsc{emproc} tasks, respectively.  Circular regions with radii of 25$^{\prime\prime}$ and 50$^{\prime\prime}$ were used for extracting source and background regions, respectively, for both pn and MOS datasets. 
Events with PATTERN $\leq$ 4 for pn and PATTERN $\leq$ 12 for MOS were selected during the reduction. The final science products were extracted with the \textsc{xmmselect}. \textsc{arfgen} and \textsc{rmfgen} tasks were used to generate the response files for each observation. The final spectral files from MOS 1 and MOS 2 from all observations constituting one epoch were merged into a single spectrum for that corresponding epoch using the ftool \textsc{addascaspec}. The net source counts for the merged pn$/$MOS spectra are provided in Table \ref{tab-1}, together with the corresponding background counts. These background values have been scaled to the source extraction region to account for the different sizes of the source and background regions\footnote{\url{https://heasarc.gsfc.nasa.gov/docs/asca/abc_backscal.html}}. All the spectra were binned to have minimum signal-to-noise (S/N) of 5 per energy bin.

\subsection{NuSTAR}
\label{sec-2.2}
In 2016, coordinated \textit{NuSTAR} \citep{2013ApJ...770..103H} observations were conducted simultaneously with the \textit{XMM-Newton} observation of PG 1535+547. \textit{NuSTAR}, equipped with focal plane modules FPMA and FPMB, captures high-energy X-ray photons ranging from 3 keV to 78 keV. \textit{NuSTAR} data were obtained from NASA's HEASARC archive\footnote{\url{https://heasarc.gsfc.nasa.gov/}} and reduced using the \textit{NuSTAR} Data Analysis Software (\textsc{nustardas}). Level-2 event files for the observations were extracted using \textsc{nuproducts}. The filters for the standard depth correction and passages through the South Atlantic Anomaly were kept at their default values: \textsc{saacalc = 3}, \textsc{saa = optimized} and \textsc{tentacle = yes}. The source photons were extracted from a circular region with a 30$^{\prime\prime}$ radius around the source. Background photons were extracted from a circular region with a 60$^{\prime\prime}$ radius, free from source photons. 
We also used the spacecraft science mode data (mode 6) following the extraction process outlined in \cite{2016ApJ...826...87W}. The final spectral files from FPMA and FPMB for Obs. 7 and 8 were merged together to a single spectrum. The spectrum is re-binned to have a S/N of 3 per energy bin. The net source and scaled background counts for the final merged spectrum are listed in Table \ref{tab-1}.

\begin{figure*}
\centering
\hspace*{-0.3cm}
\includegraphics[scale=0.45]{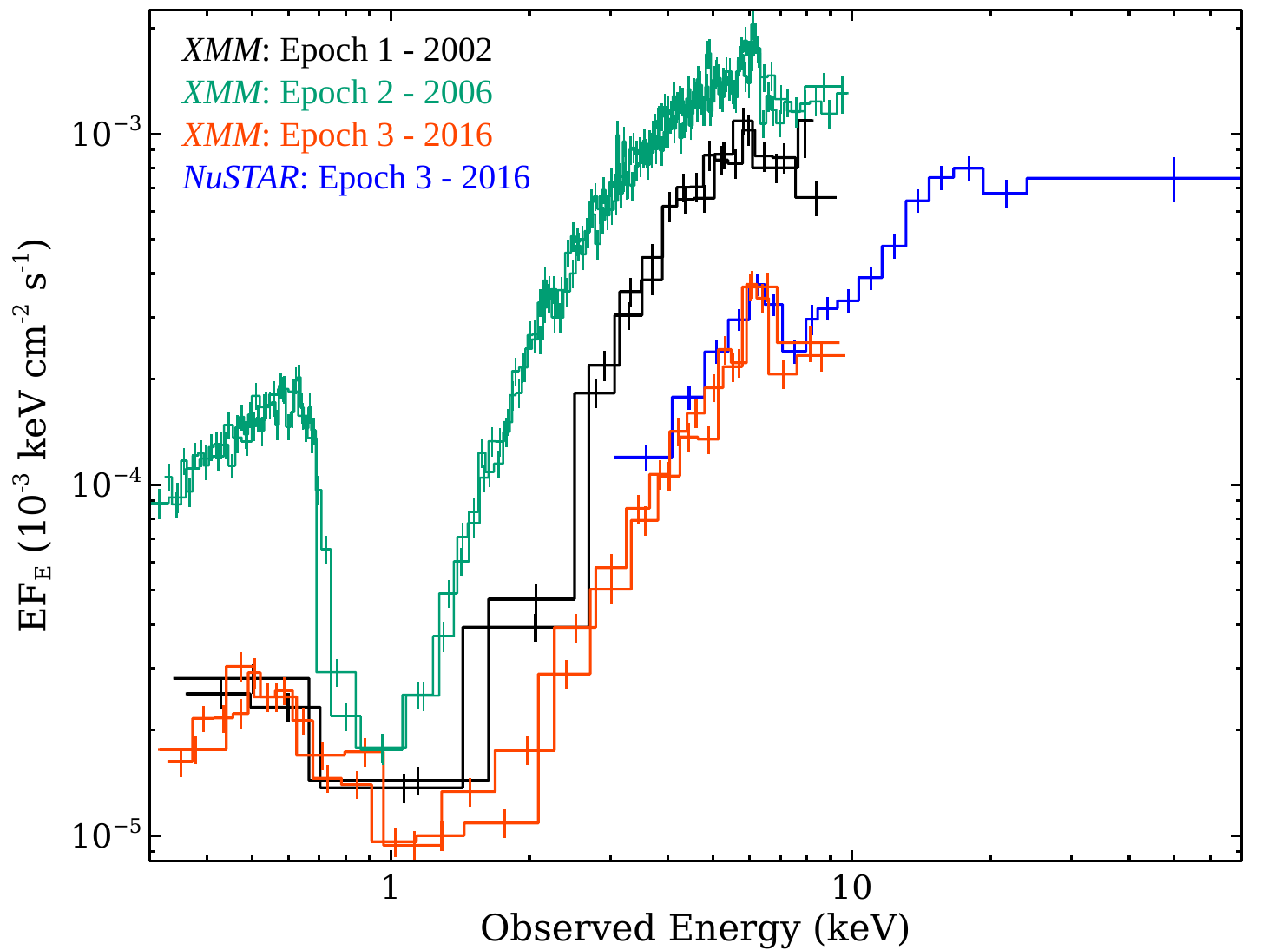}
\caption{The spectra of PG 1535+547 for all three epochs, unfolded using a simple power law model. Black and teal represent spectra from epochs 1 and 2, respectively. The orange and blue  depicts the broadband data from the coordinated \textit{XMM-Newton} and \textit{NuSTAR} observations respectively, during epoch 3.}
\label{fig-1}
\end{figure*}

\section{X-ray spectroscopy}
\label{sec-3}
\subsection{Spectral Analysis}
\label{sec-3.1}
We present a spectral analysis of all observations of the source spanning three epochs. The unfolded spectra for all three epochs are shown in Figure \ref{fig-1}. Notably, the lowest overall flux is observed from the source during epoch 3. A significant variation in spectral shape, particularly at energies $ < 3$ keV, is evident during epoch 2 compared to epochs 1 and 3. The source exhibits heightened flux and a deep broad dip between 0.7 and 2 keV, potentially indicating warm absorption from partially ionised material. This feature is likely dominated by the O VII edge at $\sim 0.7$ keV \citep{2014MNRAS.441.2613L}. In contrast, the spectra from epochs 1 and 3 do not display such absorption features below 1.5 keV, due to the enhanced levels of neutral absorption preventing us from clearly seeing the warm absorption features. Instead, these spectra exhibit a relatively steady flux between 0.3 and 1 keV. All three spectra also show variation in spectral curvature above 2 keV, with differences in flux between $2-10$ keV potentially indicating different levels of absorption across epochs. 
The steady flux below 1 keV in epochs 1 and 3, as noted earlier, suggests that additional plasma emission from larger spatial scales may be contributing in these epochs. 
The presence of emission features around $6 - 7$ keV in all three epochs indicates reflection, further confirmed by the Compton hump visible in the \textit{NuSTAR} data, peaking at $\sim 20-30$ keV. Additionally, an absorption  edge near 7 keV may be attributed to neutral absorption.

\subsection{Spectral models}
\label{sec-3.2}
Considering these spectral features, we employ a combination of models to account for various physical processes in our analysis. To describe the primary continuum and the reflection from the innermost regions of the accretion disc, we use two variants from the RELXILL \citep{2014ApJ...782...76G} family of models: \textit{relxillCp} (hereafter Model 1) and \textit{relxilllpCp} (Model 2). The former is agnostic to the X-ray source geometry and models the emissivity from the accretion disc using a broken powerlaw, while the latter explicitly defines a lamp-post geometry in which the X-ray source is modelled as an on-axis, isotropic point source at a height \textit{`h'} above the black hole. While both models ideally yield similar constraints on the reflection component, \textit{relxilllpCp} provides a direct measurement of the X-ray source height and a physically motivated interpretation of the reflection fraction $R_{\rm frac}$ \citep{2016A&A...590A..76D} by incorporating a fully specified geometry. 
We test both models to determine whether the data can only be  described by \textit{relxillCp}, which is independent of the X-ray source geometry, or if the data also supports the use of the lamppost model (\textit{relxilllpCp}), enabling us to push further and obtain constraints on the source height, and consequently, the geometry of the X-ray source itself.

The distant reflection component is modeled using the \textit{borus} table model \citep{2019RNAAS...3..173B}, where the reprocessing medium is assumed to be a sphere with conical cutouts, characterized by a half-opening angle $\theta_{\rm tor}$. Absorption due to warm material is incorporated using the photoionization table model, \textit{Xstar} \citep{2001ApJS..133..221K}. The \textit{Xstar} table grids are computed for an incident radiation of photon index $\Gamma = 2$, 
with the column density ($N_{\rm H}$) and gas ionization state (log $\xi$) of the absorbing medium left free to vary. 
Solar abundances is assumed for all elements except iron and oxygen, with the abundances of these elements included as free parameters in the table model. Turbulent velocity is set to $100~\rm km~s^{-1}$. As noted in Section \ref{sec-3.1}, obscuration due to neutral material is accounted for using the \textit{tbfeo} model \citep{2000ApJ...542..914W}, and for partial covering absorption, the respective absorbers are convolved with the \textit{partcov} model. To further address any photons that are being scattered around the absorbers into our line-of-sight  \citep{2006A&A...448..499B,2007MNRAS.374.1290G}, we add a \textit{powerlaw} model.
Additionally, to model diffuse emission from either the host galaxy or more distant regions of the AGN, we include a thermal plasma component, using the \textit{Mekal} model \citep{1985A&AS...62..197M}. We stress, though, that replacing this with an \textit{Apec} model (an alternative thermal plasma model; \citealt{2001ApJ...556L..91S}), or treating this emission instead as photoionised emission using an \textit{Xstar} grid (as suggested for other similar cases by e.g. \citealt{2023MNRAS.521.2378B,2024ApJ...971...22B}) does not impact the conclusions drawn for the key continuum/reflection/absorption parameters (e.g. $\Gamma$, $a$, $h$, $i$, $R_{\mathrm{frac}}$, $N_{\mathrm{H}}$). Note neither the scattered powerlaw emission nor the \textit{Mekal} component are absorbed by any of the nuclear absorbers.
Finally, in order to account for neutral hydrogen absorption by the interstellar medium (ISM) within our Galaxy and in the PG1535+547 galaxy, we use the \textit{tbabs} \citep{2000ApJ...542..914W} and the \textit{ztbabs} model throughout this work. The Milky Way Galactic equivalent neutral hydrogen column density, $N_{\rm H}$, along the direction of PG 1535+547 is fixed to $1.2\times10^{20} \rm cm^{-2}$ \citep{2016A&A...594A.116H}.

With these model components, we initially envisioned a relatively simple geometric configuration, where the intrinsic continuum and disc reflection from the innermost regions of the accretion disc are enshrouded by layers of ionized and neutral absorption. We treat the warm absorber (WA) seen in epoch 2 with two \textit{xstar} components, allowing one to be partially covering, and assume this absorption is still there (and is unchanged) during the other epochs in our analysis. The enhanced neutral absorption seen during epochs 1 and 3 is modeled with an additional partially covering neutral absorber that only impacts these epochs. Beyond this multi-layered obscuration, at relatively larger spatial scales, we observe contributions from distant reflection, a scattered component, and diffuse plasma emission from the host galaxy. While these model components are not affected by any of the nuclear absorbers (either warm or neutral), all model components are absorbed by the ISM absorption (both in our Galaxy and in PG1535+547).
The final best-fit model is therefore given by : $tbabs\times ztbabs(powerlaw+mekal+borus+(pcf\otimes tbfeo)\times(pcf\otimes xstar_{\rm 1})\times xstar_{\rm 2}\times relxill(cp/lpcp)$. 
We apply this model to the data from all three epochs simultaneously to fully leverage the constraints that different epochs can place on these different model components when determining the best-fit model parameters.

\subsection{Parameter Setup}
\label{sec-3.3}

While the overall model configuration remains fixed across all three epochs, we allow certain parameters to
vary to account for spectral variability. However, fundamental parameters such as black hole spin ($a$), inclination angle ($i$) of the accretion disc, and iron abundance ($A_{\rm Fe}$) are assumed invariant and therefore linked across all epochs. Parameters that are not allowed to vary between epochs, whether for physical or practical reasons, are hereafter referred to as \textit{global} parameters.

Both \textit{relxillCp} and \textit{relxilllpCp} incorporate a Comptonized primary spectrum, characterized by the photon index ($\Gamma$) and the electron temperature ($kT_{\rm e}$). We initially allow $\Gamma$ to vary across all epochs, but find it to be consistent for the two \textit{XMM-Newton}-only epochs (epochs 1 and 2), so link this parameter between them in our final analysis. 
$kT_{\rm e}$ is assumed to be the same for all epochs (though free to vary overall) as we only have one \textit{NuSTAR} observation that is sensitive to this parameter.

In \textit{relxillCp}, the emissivity profile follows a broken power-law: $\epsilon (r) \propto r^{-q_{\rm in}}$  for $r_{\rm in} \leq r \leq r_{\rm br}$ and $\epsilon (r) \propto r^{-q_{\rm out}}$  for $r_{\rm br} \leq r \leq r_{\rm out}$, where $q_{\rm in}$ and $ q_{\rm out}$ denote the inner and outer emissivity indices, while $r_{\rm in}, r_{\rm out}$ and $r_{\rm br}$ correspond to the inner, outer, and break radii of the accretion disc. The accretion disc is assumed to extend down to $r_{\rm ISCO}$ while $r_{\rm out}$ is set to $400 ~ r_{\rm g}$. Since the emissivity at $r~ >~ r_{\rm br}$ is well approximated by the Newtonian limit, we fix $ q_{\rm out} = 3$ \citep{1997ApJ...488..109R}. The parameters $r_{\rm br}$, $q_{\rm in}$ are allowed to vary across all epochs. In \textit{relxilllpCp}, the emissivity profile is instead parameterized by the X-ray source height ($h$), which is allowed to vary across all epochs, with a lower limit set at $1.5r_{\rm g}$. 
The density of the disc, log($n_{\rm e}$), is free to vary overall but is linked across all epochs. Other reflection model parameters, including reflection fraction ($R_{\rm frac}$), disc ionization (log $\xi$), and normalization, are allowed to vary across epochs. Additionally, in \textit{relxilllpCp}, we enable returning radiation and the disc ionization gradient (iongrad type = 1). Enabling the ionization gradient allows $\xi$ to vary as a power-law, $\xi \propto r^{-p}$ across the disc radius instead of assuming a constant ionization profile. The gradient index $p$ is tied across epochs.

For the \textit{borus} model component, the photon index of the ionizing continuum is set to the average of the $\Gamma$ values from the disc reflection model across all three epochs. Both $kT_{\rm e}$ and the viewing angle are directly linked to the corresponding parameters in the disc reflection model. The column density of the torus is left free to vary but remains linked across epochs, as the column density of the distant reflector is not expected to change over the observed timescales. The torus covering factor ($\Omega/4\pi$) is constrained to remain lower than the viewing angle, preventing direct viewing through the obscuring torus. While the normalization of the \textit{borus} component is not expected to vary significantly, our fits suggest a slight variation, so we allow it to vary across epochs.

\begin{table*}
\centering
\caption{Best-fitting parameters obtained from the simultaneous, broad-band analysis of
\textit{XMM–Newton + NuSTAR} data of PG 1535+547 using Model 1 (\textit{relxillCp}) and the Model 2 (\textit{relxilllpCp}): \textit{ztbabs(powerlaw+mekal+borus+(pcf$\otimes$ tbfeo)$\times$(pcf$\otimes$ xstar$_{\rm 1}$)$\times$ xstar$_{\rm 2}\times$ relxill(cp/lpcp)}. Columns 1 and 2 list the models and parameters. Columns $3 - 6$ show the results for Model 1 (global parameter values in column 3; epoch-dependent values for epochs 1, 2, and 3 in columns $4 - 6$). Columns $7 - 10$ present the corresponding results for Model 2 in the same format. See Section \ref{sec-3.3} for parameter description and assignments as global and epoch-dependent parameters. Errors are quoted at the 90\% confidence level. }

\begingroup
\setlength{\tabcolsep}{3pt}
\renewcommand{\arraystretch}{0.5}
\begin{tabular}{llcccccccc}
\hline
Model & Parameters & \multicolumn{4}{c}{\textit{relxillCp}} & \multicolumn{4}{c}{\textit{relxilllpCp}} \\
 & & \textit{global} & 2002 & 2006 & 2016 & \textit{global} & 2002 & 2006 & 2016 \\
(1) & \hspace*{0.4cm}(2) & (3) & (4) & (5) & (6) & (7) & (8) & (9) & (10) \\
\hline
\textit{ztbabs} & $N_{\rm H}$ {($ 10^{22} \rm cm^{-2}$)}  & $0.08\pm0.01$ & - & - & - & $0.09^{+0.01}_{-0.04}$ & - & - & - \\
\hline
\textit{scatt. PL}   & Norm {($10^{-6}$)}  &  $6.51^{+1.18}_{-1.64}$ & - & - & - & $7.79^{+1.03}_{-1.54}$ & - & - & - \\
\hline
\textit{mekal}  & \textit{kT}$_{\rm e}$ {(keV)}  & $0.57\pm0.05$ & - & - & - & $0.69^{+0.05}_{-0.06}$ & - & - & - \\
                & Norm ($10^{-6}$) & $3.61\pm0.07$ & - & - & - & $1.15^{+0.26}_{-0.24}$ & - & - & - \\
\hline
\textit{Borus}   & log $N_{\rm H}$ {(cm$^{-2}$)} & > 24.9 & - & - & - & $> 23.2$ & - & - & - \\
        & C factor & $0.72_{-0.05}$ & - & - & - & $0.86_{-0.02}$ & - & - & - \\
\hline

\textit{partcov} & \textit{pcf} & - & $0.81^{+0.04}_{-0.03}$ & - & $0.44^{+0.18}_{-0.16}$ & - & $0.85\pm0.07$ & - &  $0.49^{+0.13}_{-0.07}$ \\
\textit{tbfeo} & $N_{\rm H}$ {($ 10^{22} \rm cm^{-2}$)} & - & $5.07^{+1.31}_{-1.43}$ & - & $69.91^{+122.21}_{-32.73}$ & - & $2.17^{+1.02}_{-0.64}$ & - &  $59.25^{+29.0}_{-17.2}$ \\

\hline
\textit{partcov} & \textit{pcf} & $0.81^{+0.02}_{-0.04}$ & - & - & - & $0.81^{+0.02}_{-0.04}$ & - & - & - \\
\textit{xstar$_{\rm 1}$}   & $N_{\rm H}$ {($ 10^{22} \rm cm^{-2}$)} & $8.12^{+2.08}_{-0.54}$ & - & - & - & $10.17^{+0.83}_{-4.24}$ & - & - & - \\
        & log $\xi$ {(erg cm s$^{-1}$)} & < 0.42 & - & - & - & <1.90 & - & - & - \\
        & z & > 0.023 & - & - & - & > 0.032 & - & - & - \\
\hline
\textit{xstar$_{\rm 2}$}   & $N_{\rm H}$ {($ 10^{22} \rm cm^{-2}$)} &  $3.99^{+0.18}_{-0.38}$ & - & - & - & $2.53^{+0.08}_{-0.29}$ & - & - & - \\
        & log $\xi$ {(erg cm s$^{-1}$)} & $1.79^{+0.10}_{-0.04}$  & - & - & - & $1.79^{+0.02}_{-0.01}$ & - & - & - \\
        & \textit{A}$_{\rm O}$ {(solar)} & $2.0\pm 0.1$ & - & - & - & $3.2^{+0.2}_{-0.1}$ & - & - & - \\
        & \textit{z}  & $0.030^{+0.005}_{-0.009}$ & - & - & - & $0.020^{+0.003}_{-0.005}$ & - & - & - \\

\hline
\textit{Reflection} & $\Gamma$  & - & $2.30^{+0.08}_{-0.04}$$^*$ & $2.30^{+0.08}_{-0.04}$$^*$ &  $2.19^{+0.06}_{-0.30}$ &  - & $2.21^{+0.08}_{-0.05}$$^*$ & $2.21^{+0.08}_{-0.05}$$^*$ & $2.40^{+0.04}_{-0.06}$   \\
\textit{model} & log $\xi$ {(erg cm s$^{-1}$)} & - & $0.89\pm{0.39}$$^*$ & $0.89\pm{0.39}$$^*$ & < 1.34 & - &  $2.40^{+0.91}_{-0.55}$$^*$ & $2.40^{+0.91}_{-0.55}$$^*$ & < 0.69  \\
            & \textit{h} {(r$_{\rm h}$)} & - & - & - & - & - & $2.25^{+7.37}_{-0.36}$ & $6.77^{+10.24}_{-2.10}$ & < 1.72 \\
            & \textit{q}$_{\rm in}$ & - & > 6.6 & > 4.7 & > 6.4 & - & - & - & - \\
            & \textit{R}$_{\rm br}$ {(R$_{\rm ISCO}$)} & - & > 2.3 & > 1.6 & > 2.0 & - & - & - & -  \\
            & \textit{p} & - & - & - & - & $0.67^{+0.39}_{-0.30}$ & - & - & - \\
            & $R_{\rm frac}$ & - &  $0.8^{+1.4}_{-0.3}$ & $0.5^{+0.3}_{-0.2}$ & > 4.7  & - & $2.6^{+4.6}_{-2.2}$ & $0.6\pm0.3$ & $> 7.2$  \\
            & \textit{a}$^{*}$ & $0.82^{+0.13}_{-0.14}$ & - & - & - & > 0.99 & - & - & - \\
            & \textit{i} {($\deg$)} & $40.2^{+3.1}_{-4.9}$ & - & - & - & < 29.1 & - & - & - \\
            & \textit{kT}$_{\rm e}$ {(keV)} & > 16.4 & - & - & - & > 52.6 & - & - & - \\
            & log ($n_{\rm e}/\rm cm^{-3}$) & > 19.6 & - & - & - & $19.0\pm0.1$ & - & - & - \\
            & \textit{A}$_{\rm Fe}$ {(solar)} & $2.0\pm0.3$ & - & - & - &$3.1^{+0.6}_{-0.2}$ & - & - & - \\
            & Norm {($ 10^{-5}$)} & - &  $1.83^{+0.43}_{-0.40}$ & $14.11^{+7.63}_{-8.10}$ & $2.59^{+0.56}_{-0.47}$ & - & $4.50^{+0.26}_{-0.81}$  & $0.16\pm0.05$ & $9.87^{+3.35}_{-5.75}$ \\
\hline\hline
\textit{cflux}$^{\dagger}$ & log(flux) & - & $-11.54^{+0.02}_{-0.01}$ & $-11.58\pm 0.02$ & $-11.41^{+0.04}_{-0.02}$ & - & $-11.39\pm 0.01$ & $-12.25\pm 0.01$ &  $-12.08\pm 0.02$  \\
\hline
$\chi^{2}$/d.o.f & & 1407.7/1283 & - & - & - & 1402.9/1285 & - & - & - \\
\hline

\end{tabular}
\endgroup
\begin{flushleft}
     $^{*}$ indicates that corresponding parameters are linked across epochs.
     $^{\dagger}$ The \textit{cflux} estimates here correspond to the absorption-corrected flux in the $2 - 10$ keV energy band.
\end{flushleft}
\label{tab-2}
\end{table*}

The primary parameters defining the physical properties and kinematics of the warm absorbers, including the column density ($N_{\rm H}$), ionization state $\xi$, and the observed redshift ($z$), for both warm absorbers are allowed to vary during epoch 2. Additionally, the covering factor of Warm Absorber 1 (WA1) is also allowed to vary in this epoch. For Epochs 1 and 3, however, the properties of both WA1 and WA2 are linked to their corresponding values in epoch 2. Allowing the warm absorber properties to vary independently across all epochs does not offer any improvement in the fit and the WA parameters could not be constrained during epochs 1 and 3. We also tested whether removing WA1, WA2, or both during epochs 1 and 3 could still provide a good fit, but none of these configurations were successful. Therefore, we adopted a setup where WA properties are varied only in epoch 2, while those in epochs 1 and 3 are linked to epoch 2. Finally, the oxygen abundance value, primarily inferred from the oxygen line feature here, is also allowed to vary in epoch 2 and is linked across all models across all epochs.

The column density ($N_{\rm H}$) of the variable neutral obscurer, as modeled by the \textit{tbfeo} component, was initially allowed to vary across epochs. However, during epoch 2, the neutral column was found to be undetectable, with $N_{\rm H} < 10^{20} \rm cm^{-2}$. As a result, we turn off the neutral column for this epoch while allowing the partial covering factor and column densities in epochs 1 and 3 to vary. Additionally, the $\Gamma$ of the scattered component, modeled using a power-law, is set to the epoch-averaged $\Gamma$ from the disc reflection model, following the same approach as in the \textit{borus} component. With this parameter setup established, we now proceed with the spectral fitting procedures.

\section{results}
\label{sec-4}
\begin{figure}
\centering
\hspace*{-1cm} 
\includegraphics[scale=0.39]{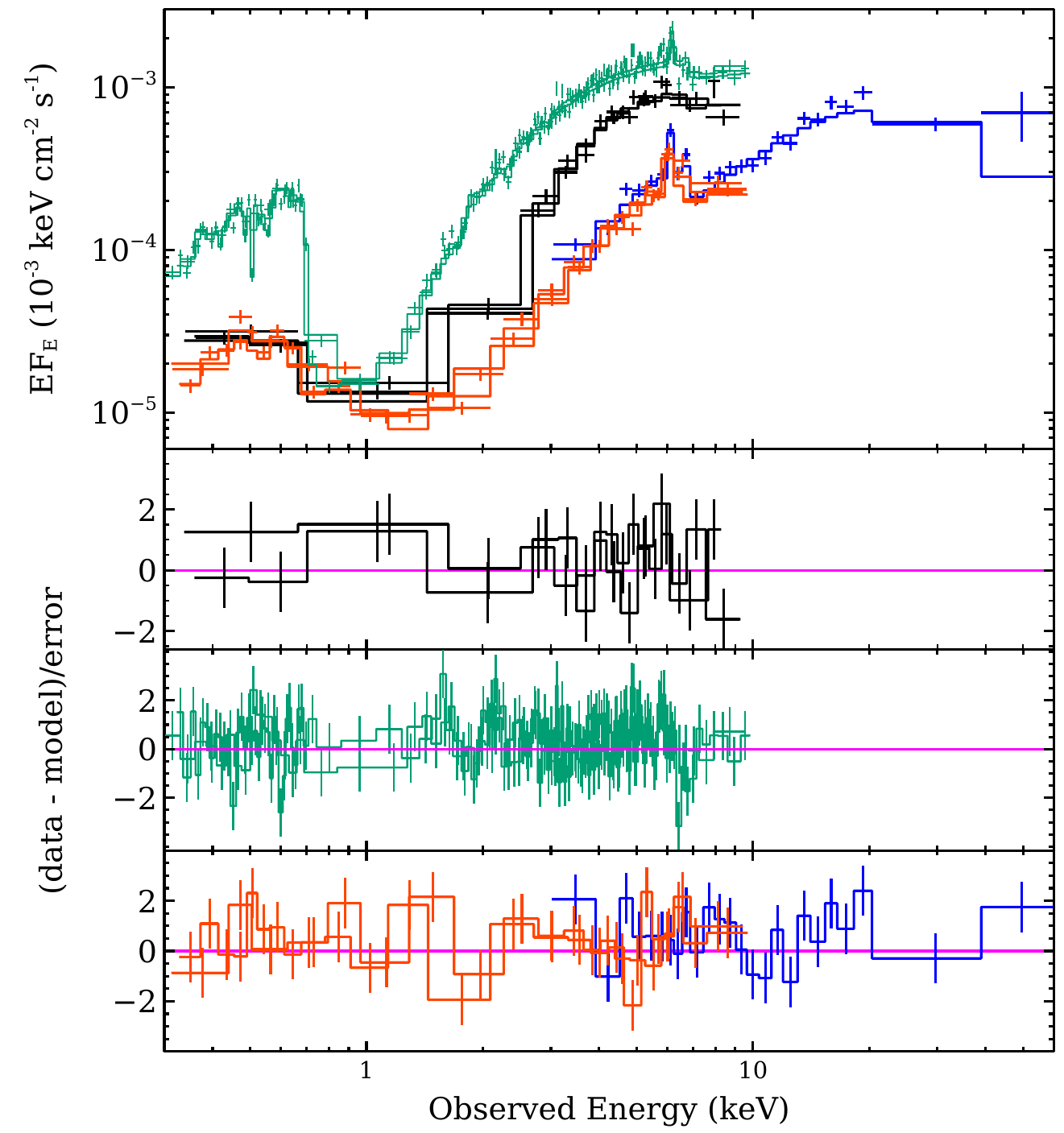}
\caption{The top panel displays the model fit obtained from the simultaneous fits for epochs 1, 2, and 3. The bottom three panels show the residuals in terms of sigma for each epoch, shown sequentially from top to bottom for epochs 1, 2, and 3. Data points are presented in the same colors as in Figure \ref{fig-1}, with the corresponding solid lines indicating the total model fit.}
\label{fig-2}
\end{figure}

\begin{figure*}
\centering
\includegraphics[scale=0.45]{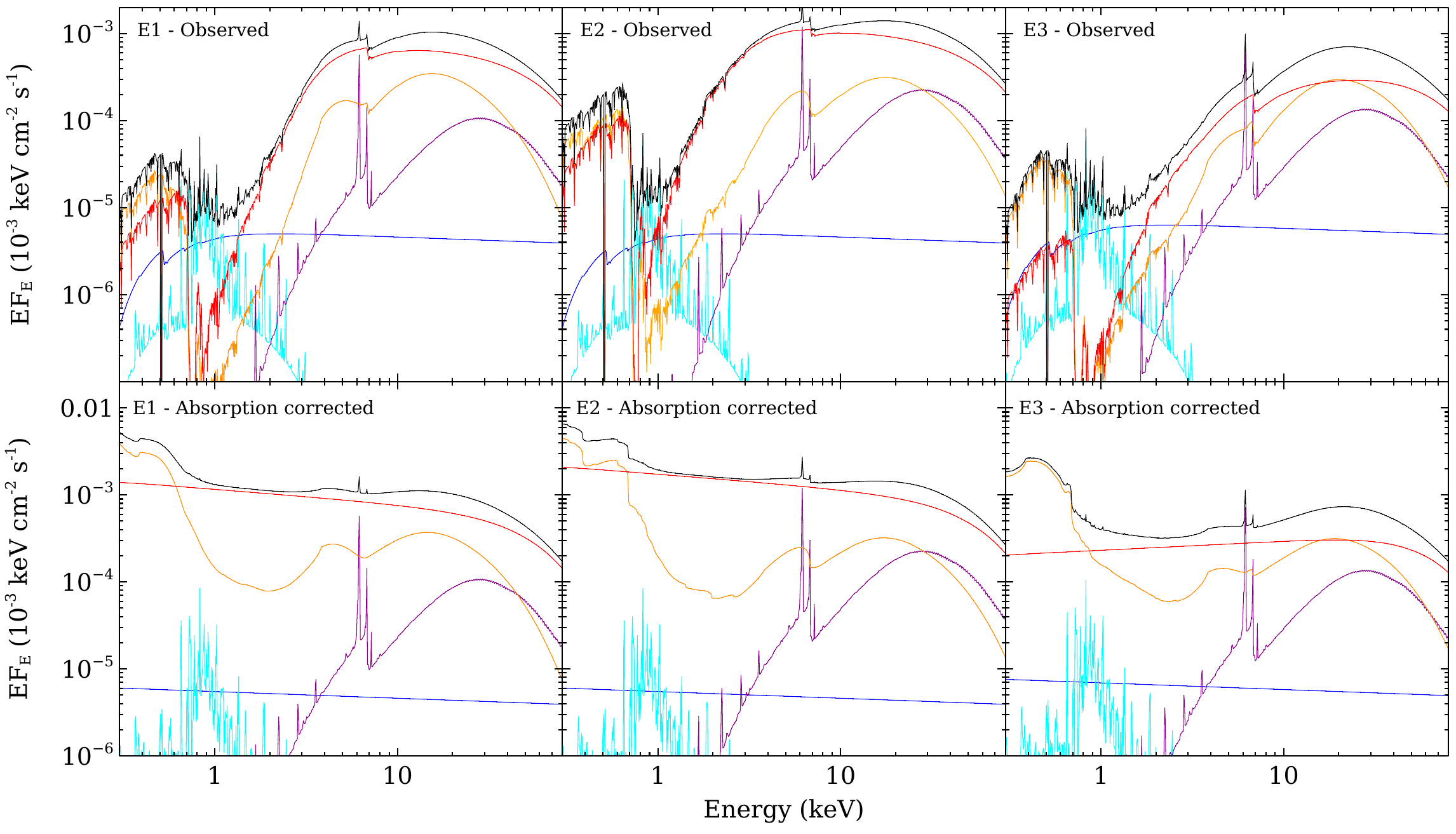}
\caption{Spectral model from Model 1, highlighting contributions from individual components to the overall spectrum of PG $1535+547$. The \textit{top} and the \textit{bottom} panels show the spectral model contributions with and without absorption, respectively. The black line shows the resultant model fitting the X-ray data, while the blue, cyan, purple, red and orange lines represent the contributions from the \textit{scattered powerlaw}, diffuse plasma emission from \textit{mekal}, distant reflection from \textit{borus}, and the {intrinsic continuum} and {inner disc reflection} components from \textit{relxillCp}, respectively.}
\label{fig-3}
\end{figure*}

\begin{figure*}
\centering
\includegraphics[scale=0.45]{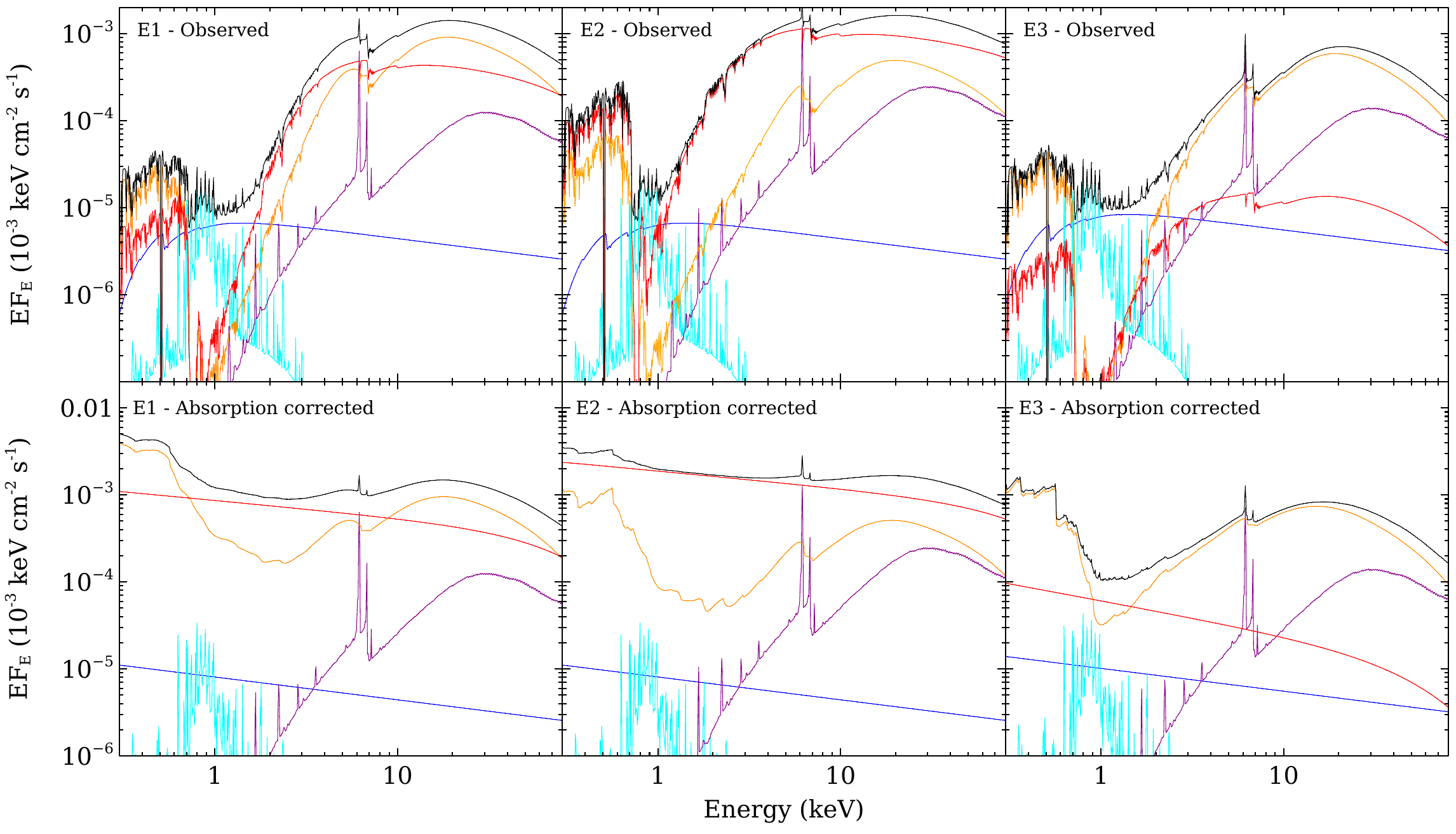}
\caption{ Spectral model from Model 2, highlighting contributions from individual components to the overall spectrum of PG $1535+547$. The \textit{top} and the \textit{bottom} panels show the spectral model contributions with and without absorption, respectively. The black line shows the resultant model fitting the X-ray data, while the blue, cyan, purple, red and orange lines represent the contributions from the \textit{scattered powerlaw}, diffuse plasma emission from \textit{mekal}, distant reflection from \textit{borus}, and the {intrinsic continuum} and {inner disc reflection} components from \textit{relxilllpCp}, respectively.}
\label{fig-4}
\end{figure*}

Models 1 and 2 produce statistically good fits with a $\chi^2/$d.o.f. of 1407.7/1283 and 1402.9/1285, respectively, each describing all features in the data comparably well and yielding broadly consistent parameters across all epochs. For illustration, the fit from Model 2 is shown in Figure \ref{fig-2} (top panel), with the corresponding residuals for epochs 1, 2, and 3 displayed sequentially in the bottom three panels. The best-fitting parameters obtained from each model for each epoch are listed in Table \ref{tab-2}. The relative contributions of the various model components for Models 1 and 2 are shown in Figures \ref{fig-3} and \ref{fig-4}, respectively. 

Our modeling confirms the initial visual indication for enhanced neutral absorption during epochs 1 and 3 (see section \ref{sec-3.1}). This absorption is strongest in epoch 3, with both models suggesting an obscurer column density, $N_{\rm H}\sim6\times10^{23}~\rm cm^{-2}$, partially covering the source with a partial covering fraction (\textit{pcf}) of  $\sim 0.5$. In epoch 1, the neutral column is lower, though still clearly present, with $N_{\rm H}\sim0.3\times10^{23}~\rm cm^{-2}$ covering $\sim 80\%$ of the emission from the central source. Epoch 2 however, shows no significant neutral absorption, with an upper limit constrained to $N_{\rm H}<10^{20}~\rm cm^{-2}$.

In addition to neutral obscuration, our fits reveal the presence of two layers of ionized absorption along the line of sight, one fully covering and one partially covering. Both models indicate the fully covering ionized absorber to have an ionization parameter of log $\xi\sim1.8$ and a column density of $N_{\rm H}\sim 3\times10^{22}~\rm cm^{-2}$. Additionally both models suggest a partially covering ionized absorber with a column density of $N_{\rm H}\sim10\times10^{22}~\rm cm^{-2}$ covering $\sim 80\%$ of the central emission.

Although variations in absorption are clearly occurring, they are not the only cause of the variations seen, as the absorption-corrected brightness of the source is also varying, as shown in the bottom panels of Figures \ref{fig-3} and \ref{fig-4}. This variation is particularly pronounced during epoch 3, where the absorption-corrected flux drops by a factor of $\sim 7$ compared to epoch 2, evident from the corresponding decline of the intrinsic continuum in the figures. Despite complex and variable absorption, the data still prefer the presence of relativistic reflection. This is evident from the significant worsening of the fit, $\Delta \chi^2 > 20$ per degree of freedom, when the disc reflection component is removed, reinforcing that a combination of complex line-of-sight absorption and relativistic reflection provides the best description of the data.

Our fits from reflection modeling suggest a $\Gamma$ value of $\sim 2.2$ for the primary continuum during epochs 1 and 2, when the source was observed exclusively with \textit{XMM-Newton}. Given their consistency, $\Gamma$ was tied between epochs 1 and 2 to reduce parameter freedom. This value is marginally steeper than those reported by \cite{2008A&A...483..137B}, who modeled the intrinsic power-law and disc reflection as passing through a partially covering ionized absorber, along with a second, low-ionization absorber fully covering the central source. For epoch 3, however, Model 2 yields a steeper continuum of $\Gamma \sim 2.4$.  The difference in $\Gamma$ values ($\Delta \Gamma \sim 0.2)$  predicted for epoch 3, by Models 1 and 2, may arise from the inclusion of returning radiation in the lamppost model (\citealt{2022MNRAS.514.3965D}), where strong gravitational light bending causes radiation from one side of the disc to curve around the black hole and illuminate the opposite side of the disc. This further enhances the reflection component, and the primary continuum steepens to compensate. 
Since returning radiation is a feature included only in the lamppost model, we observe a steeper intrinsic continuum for epoch 3 (during which light bending should be the strongest, as the inner disc is preferentially illuminated; see below) in Model 2 compared to Model 1.

\begin{figure*}
  \centering
  \begin{minipage}[b]{0.35\textwidth}
       \includegraphics[scale=0.35]{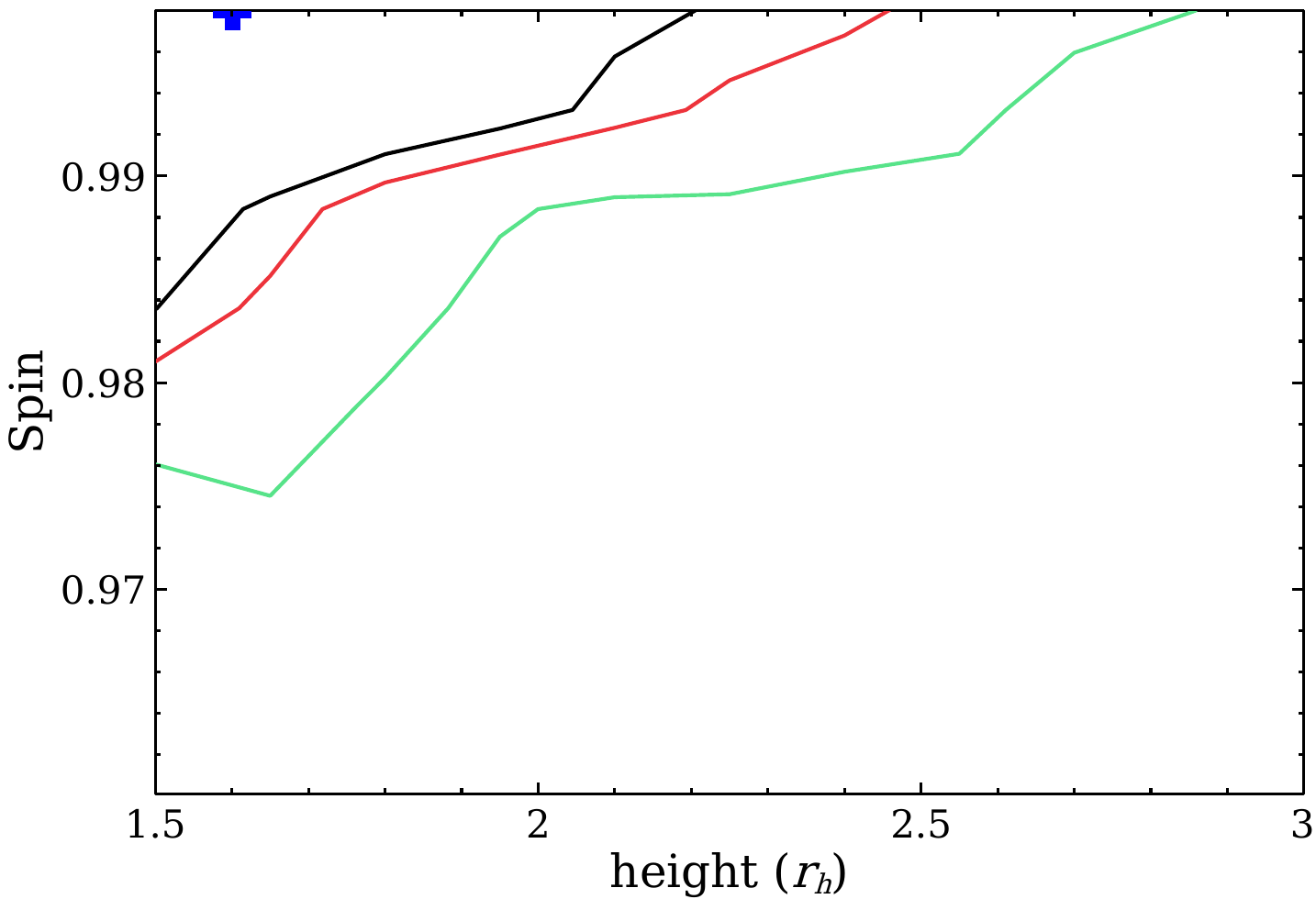}
  \end{minipage}
  \hfill
    \begin{minipage}[b]{0.35\textwidth}
 \hspace*{-2.3cm}
    \includegraphics[scale=0.35]{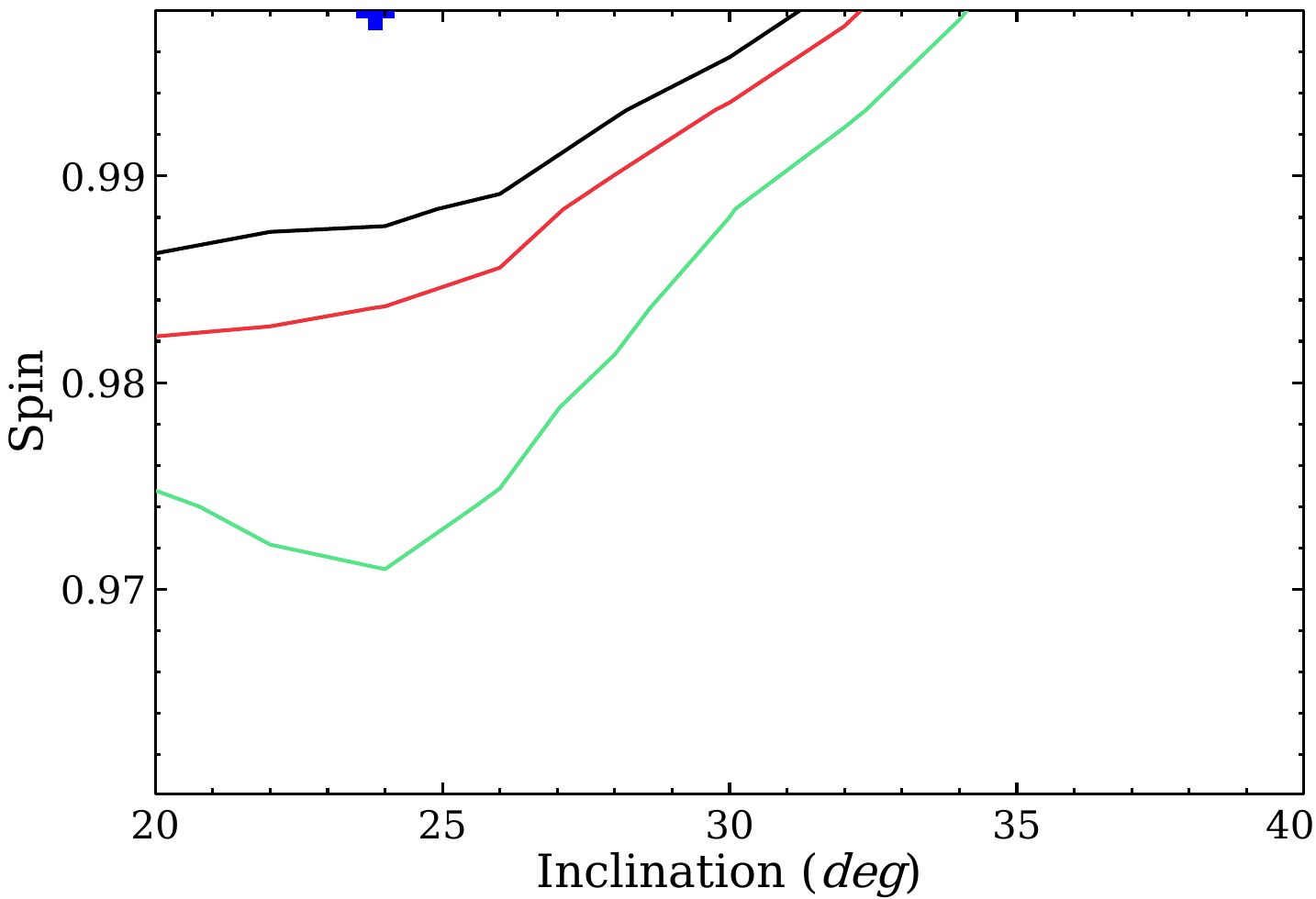}
  \end{minipage}
  
\caption{\textit{Left:} Contours for spin versus source height from the lamppost model during epoch 3, with source height expressed in unita of vertical horizon radius ($r_{\rm h}$). \textit{Right:} Contours for spin versus inclination from the lamppost model. Confidence levels are shown at 90\% (black), 95\% (red), and 99\% (green).  The blue marker in both plots indicates the best-fit values obtained from the spectral fits. \label{fig-5}}
\end{figure*}

Our fits indicate extremely strong reflection during epoch 3 as inferred from the high reflection fraction, $R_{\rm frac} > 4.7$ in Model 1 and $R_{\rm frac} > 7$ in Model 2. The lamppost model further constrains the source height to $h \leq 1.72r_{\rm g}$. The spin-height contours from the lamppost model during epoch 3, as shown in the left panel of Figure \ref{fig-5}, are in agreement with our fit results. Both models suggest a high spin value, with Model 1 yielding a spin of $0.82^{+0.13}_{-0.14}$, while Model 2 provides tighter constraints with $a > 0.99$. The tighter constraint in the lamppost case likely arises from the different treatments of the emissivity profile in the two models. In a broken power-law emissivity profile, the model allows for steep emissivity indices regardless of the actual spin (as noted in \citealt{2018MNRAS.474.1538P}), which may increase parameter degeneracies and lead to looser spin constraints. In contrast, in the lamppost geometry the emissivity profile is tied to the height of the X-ray source and the spin of the black hole, such that steep emissivity indices can only be produced by low coronal heights and rapidly rotating black holes (e.g. \citealt{2012MNRAS.424.1284W,2013MNRAS.430.1694D,2017MNRAS.472.1932G}), thereby producing much tighter constraints on the spin parameter. The enhanced reflection ($R_{\rm frac} > 7$), a high spin value $a > 0.99$ and the extreme close proximity of the X-ray source to the black hole $h \leq 1.72r_{\rm g}$, inferred from the lamppost model, indicate that the source is in a reflection dominated state during the 2016 observation. Since the effects of returning radiation become more significant for a rapidly spinning black hole and a low-height X-ray source, it makes sense that we see its influence in epoch 3, where reflection is dominant. This further supports our earlier interpretation that returning radiation could be responsible for the difference in $\Gamma$ values seen between the models.

Reflection during epoch 2 is significantly weaker than in epoch 3 but remains consistent with ``normal" levels of reflection within uncertainties ($R_{\rm frac} \sim 1$). The reflection fraction for epoch 1 remains fairly poorly constrained, particularly in the lamppost model. 
With the presence of a high level of obscuration and the absence of broadband data, the model struggles to accurately determine the strength of the disc reflection component during this epoch.

The ionization values derived across epochs 1 to 3 show slight differences between the two models. Model 1 indicates reflection from a moderately ionized disc, with log $\xi\sim 1$ at all epochs. Model 2 yields a similar ionization level during epoch 3 (with an upper limit of log $\xi$ $ < 0.7$), but a higher value in epochs 1 and 2, log $\xi\sim 2.4$, more than an order of magnitude higher than in Model 1. Model 2 includes an ionization gradient, allowing $\xi$ to vary as a power law across the disc (Section \ref{sec-3.3}), which likely explains this difference. As shown by \cite{2012A&A...545A.106S}, a compact corona can produce a radial ionization gradient across the disc, and neglecting this can lead to an overestimated emissivity index when fitting with a single ionization model. During epoch 1, when the X-ray source is inferred to be located at a height of $h \sim 2.5r_{\rm g}$, such a gradient may be expected, with ionization highest in the inner disc and declining with radius. Without accounting for this, as in Model 1, which assumes an average ionization throughout the disc, the inner disc ionization may be underestimated. To compensate, the model may then steepen the emissivity profile to mimic the reflection from a more ionized inner disc. This could explain the lower ionization values derived by Model 1 during epochs 1 and 2. The steep inner emissivity index implied by Model 1 for epoch 1, which is comparable to that seen during epoch 3 despite the lamppost fits implying the corona is located further from the black hole during epoch 1, may further support this interpretation.

Both models consistently suggest a high-density accretion disc, with log($n_{e}/\rm cm^{-3}$) > 18.9. The two models yield slightly discrepant values for the inclination angle ($i$) and iron abundance ($A_{\rm Fe}$). The inclination of the system is constrained to $i = 40^{+3}_{-5} \degree$  by Model 1, while Model 2 places an upper limit of $i < 29 \degree$. As shown in the right panel of Figure \ref{fig-5}, the spin-inclination contours from the lamppost model agree with the fit results and show no evidence of degeneracy. However, the contours indicate a broader uncertainty range for \textit{i}, with the 99\% confidence region extending closer to the uncertainty range obtained for the Model 1 fits. Similarly, the iron abundance values inferred by the two models do not formally overlap (though both are mildly super-solar), with Model 1 yielding $A_{\rm Fe} = 2.0\pm0.3$ and Model 2 yielding higher iron abundance of $A_{\rm Fe} = 3.1\pm0.2$ . These differences may stem from the slightly higher reflection fraction estimated by Model 2 compared to Model 1. This implies a stronger reflection component, including a more prominent iron line, which in turn leads to a higher inferred $A_{\rm Fe}$. 
The inclination measurement is further complicated by the presence of the neutral iron edge at $\sim 7$ keV  \citep{2015ApJ...813...84G}, which may affect the model's ability to precisely constrain this parameter. Nonetheless, it is worth emphasizing that, despite these differences, both inferred inclination angles fall within the expected range for an NLS1 galaxy, consistent with the AGN unification scheme. 

\subsection{Sanity check: Fe K emission analysis}
\label{sec-4.1}
To illustrate how the Fe K emission sits on top of the the underlying continuum after accounting for absorption effects, we re-fit the spectra for all three epochs in the $2-10$ keV energy band, adopting their respective best-fit configurations - i.e., two warm absorber components (one of which is partially covering) and, where relevant, a partially covering neutral absorber - combined with a simple powerlaw continuum instead of a relativistic reflection model. The same configuration for the warm and neutral obscurers was used to accurately reproduce the spectral curvature in this band introduced by absorption, with all absorber parameters fixed to the values determined from the simultaneous multi-epoch analysis, as many of these are primarily constrained at $E < 2$ keV and cannot be reliably determined from fits restricted to the $2 - 10$ keV band. This simpler model was first fitted over the $2 - 3.5$ keV and $7 - 10$ keV intervals, where the primary continuum should dominate, and then extrapolated to the entire $2 - 10$ keV to visualize the Fe K features.

\begin{figure}
\centering
\hspace*{-1cm} 
\includegraphics[scale=0.39]{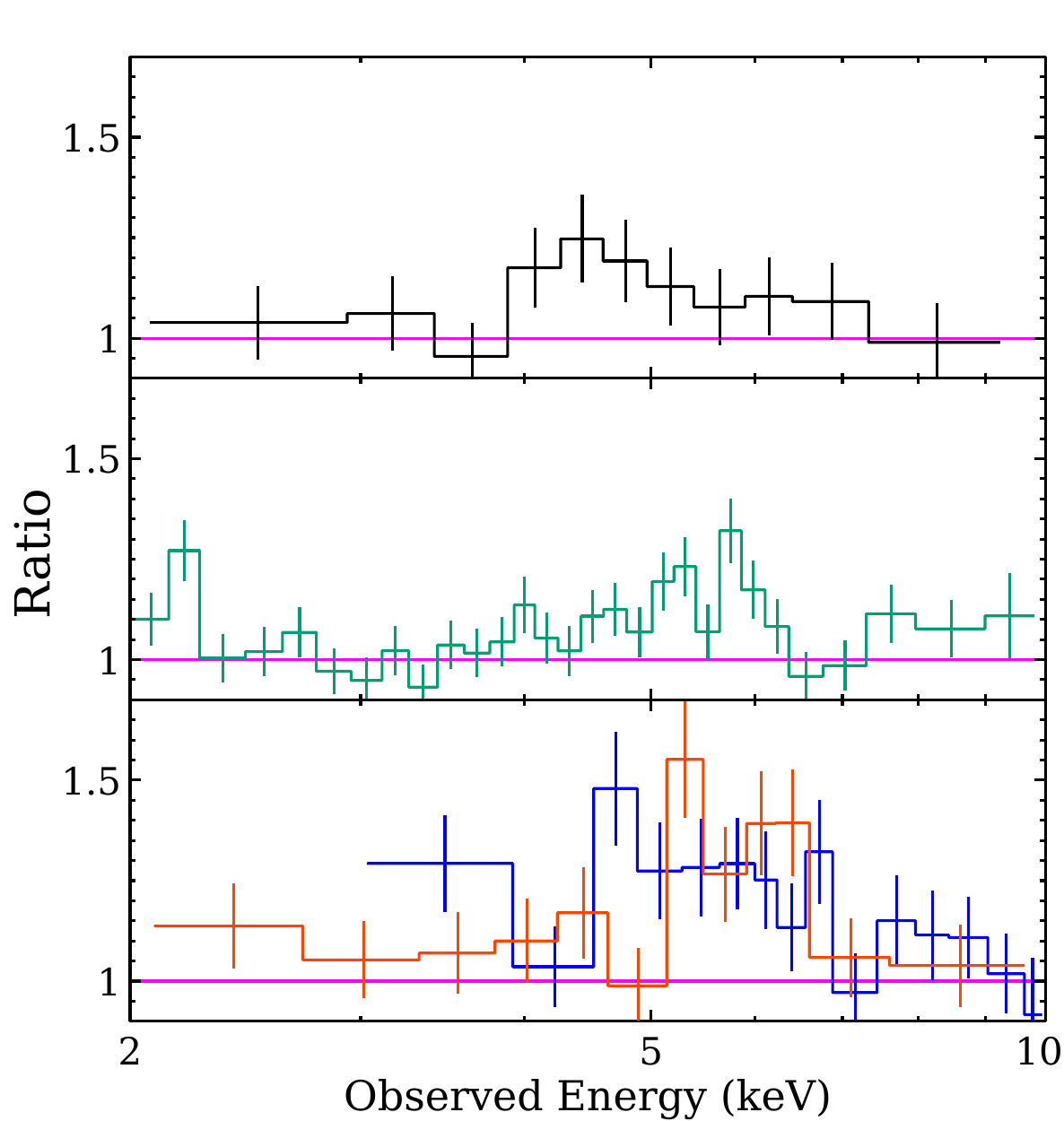}
\caption{Data-to-model ratio plots for epochs 1, 2 and 3 (\textit{top, middle} and \textit{bottom panels}, respectively) corresponding to the fits described in Section \ref{sec-4.1}. Each spectrum in the $2-10$ keV band was fitted with a simple powerlaw modified by a complex absorption geometry of the form $(pcf\otimes tbfeo)\times(pcf\otimes xstar_{\rm 1})\times xstar_{\rm 2}$,
and two emission-line components (\textit{relline} and \textit{zgauss}). The ratio plots shown correspond to the best-fit model with the \textit{relline} normalization set to zero, illustrating how the broad Fe K emission feature sits atop the absorption-corrected intrinsic continuum. Data points follow the same colour scheme as in Figure \ref{fig-1}.}
\label{fig-6}
\end{figure}

As expected, the residuals revealed both broad and narrow Fe K emission features in all epochs. The narrow emission is modeled with a Gaussian line at $\sim6.4$ keV in the source rest frame and the broad emission with a relativistic line profile using the \textit{relline} model \citep{2013MNRAS.430.1694D}. For the \textit{relline} component the line energy is restricted to $6.4 - 7$ keV in the rest frame of the AGN, corresponding to neutral and hydrogen-like iron. Assuming a single powerlaw emissivity profile of the form $\epsilon (r) \propto r^{-q}$, where $q$ is the emissivity index, we fit the $2 - 10$ keV spectra and obtain good fit with $\chi^2$/d.o.f = 87.3/90, 552/547 and 206.3/209 for Epochs 1, 2 and 3, respectively. Removing the broad line component worsens the fit by $\Delta\chi^2= 13$, 24 and 25 for the three epochs, confirming overall the presence of the broad line. The data-to-model ratio plot produced for this fit, with the \textit{relline} normalization set to zero (Figure \ref{fig-6}), illustrates the broad emission features sitting on top the underlying absorption-corrected intrinsic continuum. All epochs indicate a fairly high black hole spin, with values constrained to $a > 0.85$. 

\begin{figure}
\centering
\hspace*{-0.2cm} 
\includegraphics[scale=0.35]{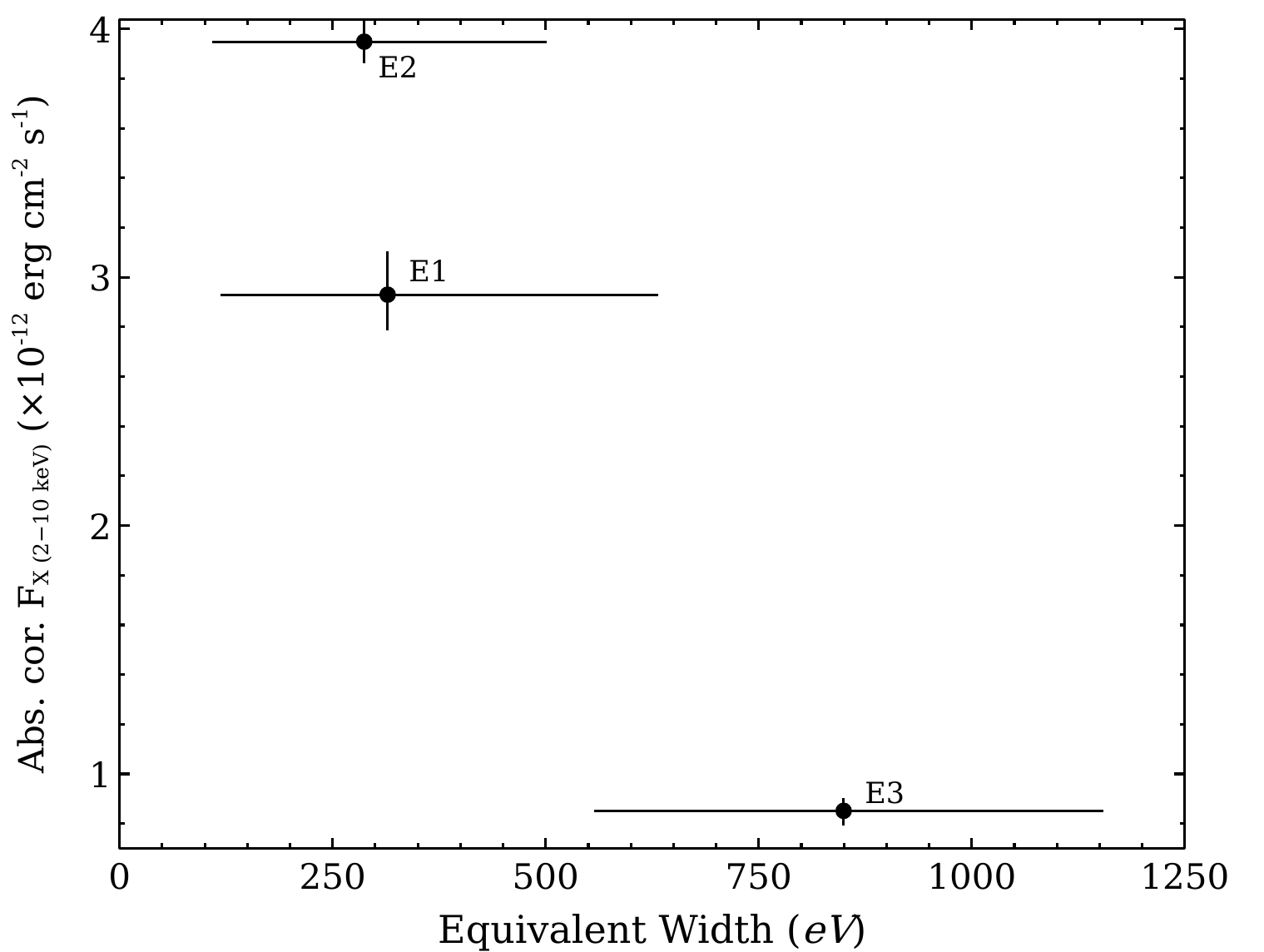}
\caption{Equivalent Width of the broad Fe K line plotted against the absorption-corrected intrinsic $2 - 10$ keV continuum flux. The continuum fluxes are derived from the fits presented in Section \ref{sec-4.1}. Points for epochs $1 - 3$ are annotated within the plot.}
\label{fig-7}
\end{figure}

The equivalent widths (EW) of the broad Fe K emission lines obtained from these fits are $EW = 314.6^{+317.2}_{-196.0}$, $280^{+214.3}_{-178.5}$ and $850^{+304.8}_{-292.4}$ eV for Epochs 1, 2 and 3, respectively. The EWs measured for epochs 1 and 2 are slightly larger than those typically expected for spectra exhibiting standard reflection with $R\sim 1$ ($\sim 180$ eV; \citealt{1991MNRAS.249..352G}), likely due to the mildly super-solar iron abundance inferred for the source (see Table \ref{tab-2}). Epochs 1 and 2 exhibit nearly identical EWs at similar intrinsic flux levels (Figure \ref{fig-7}), consistent with the roughly constant intrinsic continuum and the standard reflection regime ($R\sim 1$ within errors) shown by both epochs in the simultaneous broadband fits discussed above. Epoch 3 shows an increase in the EW together with a drop in the intrinsic flux, as seen in Figure \ref{fig-7}. This behaviour is consistent with the expectations for a reflection dominated state \citep{2004MNRAS.348.1415V,2004MNRAS.349.1435M,2010A&A...524A..50D}, in which a weaker intrinsic continuum results in an enhanced relative strength of the reflected component, producing a more prominent Fe K line. This is good agreement with the interpretation from the broadband fits, which likewise show a lower continuum flux and a higher reflection fraction ($R > 7$) during epoch 3. 

\section{discussion}
\label{sec-5}
PG 1535+547 exhibits strong spectral variability across all epochs, with its spectra well described by a combination of relativistic reflection and multiple layers of complex line-of-sight absorption. Owing to this pronounced variability and the signatures of relativistic reflection, the source presents an interesting test case for assessing the ability of relativistic reflection models to constrain the inner accretion flow properties and the black hole spin in a case involving complex and varying obscuration.

\subsection{General X-ray properties of PG 1535+547}
\label{sec-5.1}
The multi-epoch analysis of this source suggests a $\Gamma$ value varying between  2.2 and 2.4, consistent with the values typically expected for NLS1 galaxies \citep{2007ApJ...663..103L,2007MNRAS.382..194N,2017ApJS..233...17R}. The change in $\Gamma$ to a slightly steeper index ($\Gamma=2.4$) during the 2016 observation seen from model 2 still falls within the range expected for NLS1s. 
 Our estimates of $\Gamma$ is broadly consistent with previous studies \citep{2005A&A...433..455S, 2008A&A...483..137B}. Additionally, we obtained a lower limit for the electron temperature ($kT_{\rm e}$) of the source, $kT_{\rm e} > 50~ \rm keV$, marking the first time the electron temperature has been constrained for this source. Our study also infers a high solar abundance of $A_{\rm Fe} = 3.1 \pm 0.2$, which aligns with the abundance values previously estimated for this source \citep{2008A&A...483..137B}. Finally, we estimate an inclination angle of $i < 29 \degree$, which is also also in agreement with the findings by \cite{2005A&A...433..455S} and is consistent with expectations for a NLS1 galaxy based on the AGN unification scheme. This agreement serves as a robust check for our analysis.

We test PG~1535+547 for the anti-correlation between log$(n_{e})$ and log$(\dot{m}^2 m_{\rm BH})$, as proposed by \cite{1994ApJ...436..599S}, where the electron density in a radiation-pressure-dominated disc is given by  
\begin{equation}
   n_e = \frac{1}{\sigma_{T}}\frac{256\sqrt{2}}{27}\alpha^{-1}R^{3/2}\dot{m}^{-2}[1-(R_{\rm in}/R)^{1/2}]^{-2}\xi^\prime(1-f)^{-3},
   \label{eq-1}
\end{equation}
where $\sigma_{T} = 6.64\times10^{-25}$ cm$^{2}$ is the Thomson cross-section; $\dot{m}$ is the dimensionless accretion rate; $R_{\rm in}$ is the inner disc radius (set to 2$R_{\rm g}$); $R_{\rm s}$ is the Schwarzschild radius; \textit{R} is the disc radius in units of  $R_{\rm s}$; and $\alpha$ is the disc viscosity parameter (set to 0.1).  $\xi^\prime$ is the conversion factor in the radiative diffusion equation and is set to 1. The quantity \textit{f}, denoting the fraction of disc power transferred to the corona, varies between 0 and 1, with $f=0$ reducing to the standard thin disc solution of \citealt{1973A&A....24..337S}. 

\begin{figure}
\centering
\hspace*{-0.3cm}
\includegraphics [scale=0.37]{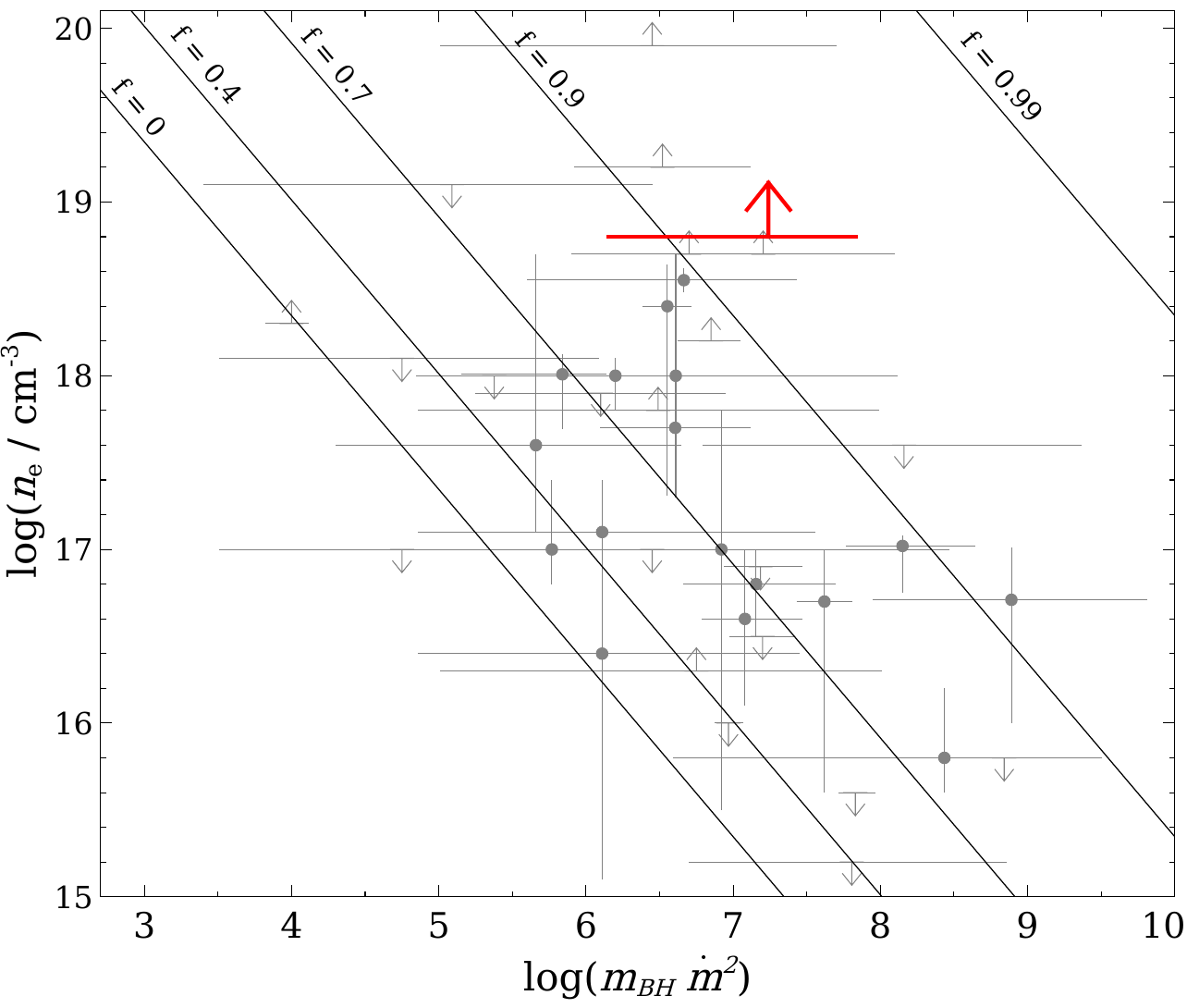}
\caption{Plot of log ($n_{\rm e}/\rm cm^{-3}$) vs log ($\dot{m}^2 m_{\rm BH}$) showing PG $1535+547$ (red point) in comparison with a sample of 31 Sy 1 AGN (black points in the background), compiled from \protect\cite{2018MNRAS.479..615M,2019ApJ...871...88G,2019MNRAS.489.3436J,2021MNRAS.506.1557W,2021ApJ...913...13X,2022MNRAS.513.4361M,2022MNRAS.514.1107J,2024MNRAS.534..608M} and \protect\cite{2025MNRAS.543.2633W}. Black lines denote solutions of the density of the radiation pressure dominated accretion discs with \textit{f} values of $0, 0.4, 0.7, 0.9$ and $0.99$, as indicated. See text for details. }

\label{fig-8}
\end{figure}

The dimensionless mass accretion rate, $\dot{m}$, is estimated using the relation: $\dot{m} = L_{\rm Bol}/\eta L_{\rm Edd}$, where $L_{\rm Bol}$ and $L_{\rm Edd}$ are the bolometric and Eddington luminosities and $\eta$ is the radiative efficiency. We assume $\eta= 0.3$, owing to the rapid spin value estimated from our analysis. The monochromatic luminosity of the source at 5100~A$\degree$ is estimated to be $\lambda L_{\lambda}\sim0.7\times10^{44}~\rm {ergs~s^{-1}}$ \citep{2021ApJS..253...20H}, from which we derive $L_{\rm Bol}$ using the scaling relation: $L_{\rm Bol}= 9\lambda L_{\lambda}(5100~A\degree)$. For a black hole mass of $15.5 \times 10^{6} M_{\odot}$ and $\dot{m} = 1.05$, we calculate log$(\dot{m}^2 m_{\rm BH}) \sim 7.2$. PG 1535+547 is plotted as the red point in Figure \ref{fig-8}, which displays a sample of Type 1 AGN compiled from \cite{2018MNRAS.479..615M}, \cite{2019ApJ...871...88G}, \cite{2019MNRAS.489.3436J}, \cite{2021MNRAS.506.1557W},  \cite{2021ApJ...913...13X}, \cite{2022MNRAS.513.4361M}, \cite{2022MNRAS.514.1107J}, \cite{2024MNRAS.534..608M} and \cite{2025MNRAS.543.2633W}. The source lies well within the distribution of the comparison sample and sits above the theoretical solution for the coronal dissipation fraction of $f=0.7$, indicating that more than 70\% of the disc's accretion power is dissipated into the corona.


\subsection{Spin of the black hole in PG 1535+547}
\label{sec-5.2}
One of the primary goals of this study is to determine whether current relativistic reflection models can effectively constrain the spin of PG 1535+547 using reflection spectroscopy, despite the challenges posed by obscuration. By employing a high-disc-density relativistic reflection model in conjunction with broadband data, we are able to model the reflection from the innermost regions, trace the imprint of gravitational effects on the reflection component and derive constraints on the spin of the black hole. The two models considered here both imply the presence of a rapidly rotating black hole, though formally give different quantitative constraints: the lamppost model gives a tighter one-sided constraint of $a > 0.99$ and the broken powerlaw emissivity model gives a looser constraint of $a = 0.82^{+0.13}_{-0.14}$. However, as discussed by several authors previously, the steep emissivity indices found in the broken powerlaw emissivity model require compact coronae around rapidly rotating black holes (e.g. \citealt{2012MNRAS.424.1284W,2013MNRAS.430.1694D,2017MNRAS.472.1932G}). As noted above, the tighter spin constraint in the lamppost model likely comes from the fact that this requirement is explicitly included in the reflection model, while it is not encoded in the broken powerlaw emissivity model. As such, we adopt the constraint of $a > 0.99$ returned by the lamppost model as our preferred spin constraint for PG 1535+547.

While previous studies by \cite{2005A&A...433..455S} and \cite{2008A&A...483..137B} attempted to model the reflection component using early relativistic reflection models (\textit{kdblur}; \citealt{1991ApJ...376...90L} and \textit{Reflion}; \citealt{1999MNRAS.306..461R}) based on \textit{XMM-Newton} observations, a key advantage of this study is the broadband coverage leveraged here, which was not available for those prior works. The extended energy range provided by \textit{XMM-Newton} and \textit{NuSTAR} enables a more reliable spectral decomposition than \textit{XMM-Newton} alone, leading to improved constraints on the reflection parameters.


In PG 1535+547, we observe complex obscuration, including a fully and partially covering warm absorber, as well as a partially covering neutral absorber. It is well established that the presence of complex absorption can result in it being challenging to identify and characterize the relativistic reflection from the inner disc, particularly in the absence of broadband spectral coverage. 
Similarly, in our case, the flux dimming observed in 2002 and 2016 could have been solely attributed to variability in the obscuring medium. However,  thanks to \textit{NuSTAR} data extending to broadband energies, combined with the multi-epoch data probing different levels of obscuration towards the source, we are able to disentangle the effects of obscuration and reflection. This allows us to accurately model the X-ray continuum up to hard X-ray energies, precisely characterize the reflection component, and place reliable constraints on the spin. The ability to constrain spin using broadband data despite complex obscuration has also been demonstrated in previous studies, e.g., \cite{2018MNRAS.480.3689B, 2018MNRAS.473.4377W, 2019ApJ...871...88G, 2020MNRAS.498L.140P, 2020MNRAS.499.1480W}.

The importance of high S/N broadband data in constraining black hole spin using current reflection models has previously been emphasized in \cite{2016MNRAS.458.1927B} and \cite{2018A&A...614A..44K}. These authors simulated broadband \textit{XMM-Newton} and \textit{NuSTAR} spectra for Seyfert 1 AGN, incorporating multiple components commonly observed in AGN, such as warm and partially covering neutral absorbers, relativistic and distant reflection along with the intrinsic continuum. Their findings demonstrated that high S/N broadband data could reliably constrain black hole spin, even in the presence of complex obscuration. Their findings highlighted the critical role of the height of the X-ray source in determining the reflection contribution, particularly for sources with spin parameter $\geq0.8$ and a lamppost height $\leq 5r_{\rm g}$. They also noted challenges in constraining spin for sources with higher lamppost heights, which sometimes led to inaccurate spin estimates. In this context, the coordinated \textit{XMM-Newton + NuSTAR} observation in 2016 provided a crucial contribution, capturing PG 1535+547 in a reflection-dominated state. During this epoch, the source was detected at a lower height ($h\leq 1.72r_{\rm g}$), enabling us to constrain its spin in this study. In contrast, during the other two epochs, observed only with \textit{XMM-Newton}, our reflection model inferred higher heights, which could have presented challenges in constraining the spin if only these data had been available.  

Placing these spin estimates in a cosmological context, our results indicating a rapid rotation of the black hole in PG 1535 ($a > 0.99$) support a scenario of steady, radiatively-efficient accretion history for the black hole in PG 1535+547. This finding aligns well with predictions from semi-analytic models \citep{2014ApJ...794..104S,2019ApJ...873..101Z} and hydrodynamic simulations \citep{2019MNRAS.490.4133B} for the growth of SMBHs with masses similar to that of PG 1535+547 (see also \citealt{2024FrASS..1124796P,2025arXiv250115380M}).

\subsection{Light bending in PG 1535+547 during the reflection dominated state}
\label{sec-5.3}
Our study reveals significant variability in the X-ray source across all three epochs. During the least obscured phase, the source appeared consistent with ``normal" reflection ($R_{\rm frac} \sim 1$) within errors, and a source height of $h \sim 7r_{\rm g}$. However, in the most obscured state, observed in 2016, the source exhibited an exceptionally high reflection fraction ($R > 7$), with the X-ray emitting region located extremely close to the black hole ($h \leq 1.72r_{\rm g}$). This was accompanied by a drastic decrease in intrinsic flux, dropping by a factor of $\sim 7$ in the $2-10$ keV band, between 2006 and 2016.

This sharp decline in X-ray flux, along with the changes in reflection characteristics, is consistent with the light-bending phenomenon proposed by \cite{2004MNRAS.349.1435M}. They demonstrated that when the height of the X-ray source falls below $2r_{\rm g}$, only a small fraction of photons escape to infinity, while most are either redirected toward the accretion disc or lost into the black hole (for an illustrative schematic of this compact corona, light bending geometry, see Figure 12 of \citealt{2012MNRAS.424.1284W}). This leads to an enhanced reflection fraction and a significant suppression of the directly observed X-ray flux. A few other well-known AGN that have exhibited strong light bending effects include 1H 0707--495 \citep{2012MNRAS.419..116F}, IRAS 13224--3809 \citep{2013MNRAS.429.2917F,2015MNRAS.446..759C}, ESO 033--G002 \citep{2021MNRAS.506.1557W,2025arXiv251013337N} and Mrk 335 \citep{2014MNRAS.443.1723P}.

In PG 1535+547, the variability observed during the 2016 epoch appears to be driven primarily by changes in the Comptonized continuum, as illustrated in Figures \ref{fig-3} and \ref{fig-4}. While the reflection component does show some variability across epochs, its changes are modest compared to those in the continuum. This behaviour is consistent with the reflection-dominated low-flux states observed in MCG--6--30--5 \citep{2007PASJ...59S.315M}, IRAS 13224--3809 \citep{2013MNRAS.430.1408K}, and Mrk 335 \citep{2014MNRAS.443.1723P}, where, under the light-bending framework, such variability is attributed mainly to coronal height changes affecting the intrinsic continuum, with the reflection component exhibiting relatively lower variability \citep{2004MNRAS.349.1435M}.

\begin{figure}
\centering
\hspace*{-0.3cm}
\includegraphics [scale=0.3]{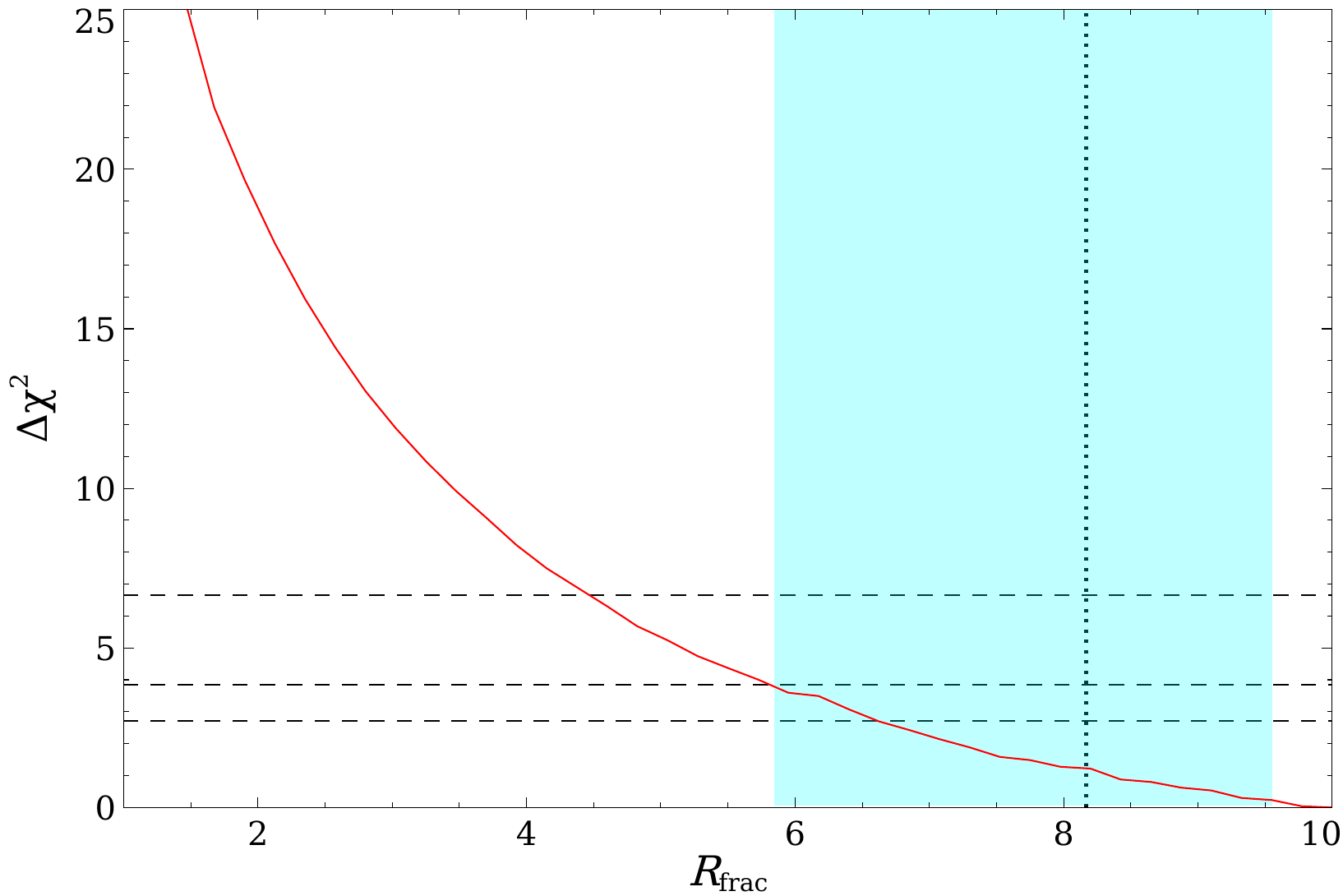}
\caption{The $\Delta\chi^2$ contour for $R_{\rm frac}$ based on the spectral fit for epoch 3 using Model 2, is shown as the solid red line. The vertical dotted line marks the predicted $R_{\rm frac}$ for this epoch, computed self-consistently from the measured spin and coronal height within the lamppost geometry. The shaded region indicates the 90\% uncertainty range on the predicted value, and the dotted lines denote the 90\%, 95\%, and 99\% confidence intervals. The predicted and observed values are in excellent agreement.}
\label{fig-9}
\end{figure}

Reflection-dominated states such as that observed here require both a compact corona and a rapidly rotating black hole in order to maximize the amount of coronal radiation that illuminates the inner disc (see, e.g., \citealt{2014MNRAS.444L.100D}). The high spin inferred from our spectral fits ($a > 0.99$), independently determined from the profile of the reflected emission, is fully consistent with this scenario. Furthermore, the predicted reflection fraction for epoch 3, computed self-consistently within the lamppost geometry from the measured spin and coronal height ($R_{\rm frac} \sim 8.2$), closely matches the observed value for this epoch ($R_{\rm frac} > 7$), as shown in Figure~\ref{fig-9}. This agreement, together with extremely low coronal height ($h \leq 1.72r_{\rm g}$), provides strong support for a compact, X-ray source located very close to the black hole during this epoch. 

We stress that invoking such a geometry in this case is not inconsistent with the recent \textit{IXPE} results that have shown a preference for slab-like coronal geometries over lamppost models in other AGN (e.g. \citealt{2023MNRAS.525.5437I,2023MNRAS.523.4468G}). None of these \textit{IXPE} results probe reflection-dominated states comparable to that seen during epoch 3, where the compact, lamppost-like coronal geometry is required. Instead they all probe relatively modest reflection fractions, more comparable to our results from epochs 1 and 2 where a `normal' reflection fraction (consistent with $R \sim 1$) is seen. Although in model 2 the coronal geometry is still nominally treated as a lamppost during these epochs, we note that its main geometric parameter (the height) is relatively weakly constrained in both cases, and that the `normal' reflection fractions can easily be reproduced by a variety of coronal geometries. We could therefore easily imagine a scenario in which the reflection-dominated state seen during epoch 3 has come about because of a radial contraction of a more slab-like geometry (as suggested by \textit{IXPE} for other AGN) during epochs 1/2 into something compact and lamppost like during epoch 3.

Further support for the light-bending interpretation in other NLS1s come from X-ray reverberation lag studies, where time delays between variations in the direct continuum and the reflected emission can probe distances down to tens of lightseconds from the black hole. In IRAS~13224-3809, for instance, \cite{2013MNRAS.430.1408K} detect shorter reverberation lags when the source is seen to have a lower flux, a result further confirmed by
\cite{2020NatAs...4..597A}. 
This suggests a shorter light-crossing time during low-flux intervals, consistent with a more compact X-ray source located closer to the black hole, irradiating only the innermost disc regions due to strong light bending. 
Unfortunately, the presence of the complex absorption and the lower S/N of the data available here mean similar reverberation studies are not yet feasible for PG~1535+547.

It is worth noting that another geometric factor may potentially also contribute to the high reflection fraction in epoch 3. In high accretion regimes such as that of PG~1535+547, the inner disc may become puffed up, forming a slim disc that deviates from the standard thin-disc approximation (e.g. \citealt{1973A&A....24..337S,1988ApJ...332..646A}), creating more of a funnel/bowl-like geometry for the innermost regions. Such a change in geometry could enhance the observed reflection fraction if the corona were located (at least mostly) within this inner bowl, allowing the inflated disc to intercept and reprocess a larger fraction of coronal photons. 
However, we note again that we also observe ``normal'' reflection fractions (consistent with $R \sim 1$) during epochs 1 and 2, when the absorption-corrected flux was much brighter than the reflection-dominated epoch 3. This would imply that such a geometry is not relevant during these epochs, either because the disc is not particularly thick or because the corona is not located within the inner bowl. Either way, the reflection-dominated state seen during epoch 3 still implies a contracting of the corona must have occurred. Given then the coronal height of $\sim7r_{\rm g}$ inferred for epoch 2, should such a bowl-like geometry represent the inner flow it is likely that in order to place the corona $\sim$fully within this bowl it would need to be sufficiently compact and close to the black hole that strong relativistic light bending would be a relevant factor anyway. Nevertheless, it is possible that such a geometry could help further enhance the observed reflection fraction seen during this epoch in addition to strong light bending. Reflection models that self-consistently allow for such a geometry (e.g. \citealt{2018ApJ...868..109T}) will be required to test this possibility in the future. We also note that the reflection fraction does not directly influence the spin parameter in the fits presented here, and that our spin constraint is based on a simultaneous fit to all epochs, so this constraint is likely robust to such issues.



\subsection{Absorbers in PG 1535+547}
\label{sec-5.4}
PG 1535+547 exhibits a complex, multi-layered absorption along the line of sight, characterized by two warm absorbers (WAs): one partially covering (WA1) and the other fully covering (WA2) the intrinsic emission. We also see the presence of a partially covering neutral absorber, during 2002 and 2016.

The partially covering warm absorber (WA1) has a high column density, with a column density of $N_{\rm H}\sim1\times10^{23}~{\rm cm^{-2}}$
and an ionization parameter of log $\xi$ (erg cm s$^{-1}$) $ < 1.9$, partially covering 80\% of the intrinsic emission. The fully covering warm absorber (WA2) exhibits a slightly lower column density, 
$N_{\rm H}\sim3\times10^{22}~{\rm cm^{-2}}$, but a similar ionization state to WA1. While WA1 does not show any evident Doppler shift (we constrained an upper limit on its outflow velocity to be $v_{\rm WA1} < 1450~\rm km~s^{-1}$), WA2 is found to be outflowing with a velocity of $v_{\rm WA2}\sim 5000~{\rm km~s^{-1}}$. While these column density and velocity parameters of WA1 and WA2, for the derived ionization parameter (log $\xi\sim1.8$), are slightly higher than those observed in the Warm Absorbers in X-rays (WAX) sample \citep{2014MNRAS.441.2613L}, they are well within the extended X-Ray Winds in Nearby-to-distant Galaxies (X-WING) sample, as compiled by \cite{2024ApJS..274....8Y}. 
Additional neutral obscurers with column, $N_{\rm H}\sim3\times10^{22}~{\rm cm^{-2}}$ and $6\times10^{23}~{\rm cm^{-2}}$ are seen obstructing the source during 2002 and 2016, respectively, covering 80\% and 50\% of the source emission. 

Based on the column density, ionization state and outflow velocities, and drawing analogies to previously studied X-ray warm absorber obscured Sy1s, we propose a stratified absorption configuration for obscuration in PG 1535+547. WA1 is conceptualized as an ionized disc wind outflowing from the accretion disc, similar to what has been observed in, e.g., NGC 3516 \citep{2008A&A...483..161T}, NGC 3783 \citep{2017A&A...607A..28M}, and NGC 3227 \citep{2015A&A...584A..82B}.  Meanwhile, the fully covering WA2 could be associated with clumpy structures within the broad-line region (BLR) or even the more distant narrow-line region (NLR), as seen in, e.g., NGC 5548 \citep{2014Sci...345...64K,2015A&A...575A..22M}, UGC 3142 and ESO 140-43 \citep{2010A&A...518A..47R}.

The varying column densities and covering fractions of the neutral obscurer, observed in 2002 and 2016 observations, may point to the presence of inhomogeneous, neutral material at the inner edge of the torus \citep{2009ApJ...695..781B,2014MNRAS.437.1776M}. Alternatively, it might also resemble the transient obscuration seen in NGC 5548, where the obscuring material lies interior to the WAs (see Figure 4 in \cite{2014Sci...345...64K} and \cite{2022ApJ...934L..24M} for an illustration of one of the potential absorber configurations). In that system, the obscuring wind shields the WA from the ionizing continuum, leading to a drop in its ionization state. In our case, however, since we have had to link the WA properties across epochs (see Section \ref{sec-3.3}), we have not been able to directly track any WA variability in response to changes in the neutral obscurer. As a result, our work do not allow us to draw firm conclusions about the location of the obscurer. 

Following the approach outlined in \cite{2012MNRAS.422L...1T}, we can constrain the location of the warm absorbers using the equation: $r \leq r_{\rm max} = L_{\rm ion}/\xi N_{\rm H}$,  where $r_{\rm max}$ is the maximum distance of the gas from the X-ray source, $L_{\rm ion}$ is the unabsorbed ionizing luminosity between 13.6 eV to 13.6 keV, $\xi$ is the ionization parameter and $\rm N_{\rm H}$ is the absorber column density. Additionally, we estimate the minimum distance of the absorbers by equating their outflow velocities to the escape velocity: $r \geq r_{\rm min} = 2GM_{\rm BH}/v_{\rm out}^2$. Using these constraints, we estimate WA1 to be located within the range  $10^{4}r_{\rm s} \leq r_{\rm WA1} \leq 10^{6}r_{\rm s}$, and WA2 to be located anywhere between $10^{3}r_{\rm s} \leq r_{\rm WA2} \leq 10^{7}r_{\rm s}$. The ranges estimated above loosely places the WAs spanning anywhere from the outer accretion disc, the edge of the BLR region or even the NLR region. 


\section{conclusions}
\label{sec-6}
We performed a multi-epoch X-ray reflection spectroscopy analysis of PG 1535+547 using \textit{XMM-Newton} observations from 2002 and 2006, along with a coordinated \textit{XMM-Newton + NuSTAR} observation in 2016, covering a broad energy range of $0.3 - 60$ keV. The source exhibits significant variability across all three epochs. Our analysis employs a combination of relativistic reflection modeling and multiple layers of complex absorption along the line-of-sight. Thanks to the broadband coverage provided by \textit{XMM-Newton} and \textit{NuSTAR}, we were able to place robust constraints on several key parameters of the source, including black hole spin, despite the challenges posed by the complex obscuration along our line of sight. Our key results are:

\begin{enumerate}
    \item The intrinsic spectrum of the source is characterized by a primary continuum with $\Gamma \sim 2.2 - 2.4$, and the electron temperature is constrained to $kT_{\rm e} > 52$ keV. We infer a slightly elevated iron abundance of $A_{\rm Fe} = 3.1$, and our fits suggest an upper limit on the inclination angle, indicating that the source is viewed at an angle of $i < 29\degree$.
    \item Our fits indicate that the black hole in PG 1535+547 is rapidly rotating, with a spin value of $a > 0.99$. 
    \item In 2016, the source was observed in a reflection-dominated state, exhibiting extreme reflection ($R_{\rm frac} > 7$) and an X-ray emitting region located very close to the black hole ($h \leq 1.72r_{\rm g}$). The accompanying flux drop by a factor of $\sim7$ supports the light-bending model. 
    \item Our analysis suggests that the source is surrounded by three layers of absorption: \textit{(i)} a partially covering warm absorber (WA1) with $N_{\rm H}\sim1\times10^{23}~{\rm cm^{-2}}$ and  ionization parameter with log\,~$\xi$\,(erg\,~cm\,~s$^{-1}$)\,$<1.9$, covering $\sim80\%$ of the intrinsic emission; \textit{(ii)} a fully covering warm absorber (WA2) with $N_{\rm H}\sim3\times10^{22}~{\rm cm^{-2}}$ and similar ionization as WA1; and \textit{(iii)} a partially covering neutral obscurer, detected only in the 2002 and 2016 observations, with varying column densities and covering fractions.  Due to the lack of continuous or multi-wavelength observations, we are unable to precisely determine the locations of these absorbers. However, based on our X-ray estimates, we infer that they could reside anywhere within $10^3-10^7 r_{\rm s}$, consistent with regions spanning from the outer accretion disc to the more distant narrow-line region.
\end{enumerate}

\section*{Acknowledgements}
We thank the anonymous referee for their helpful comments and suggestion. DJW acknowledges support from the Science and Technology Facilities
Council (STFC; grant code ST/Y001060/1). TD acknowledges support from the DFG research unit FOR 5195 (grant number WI 1860/20-1).  The work of DS was carried out at the Jet Propulsion Laboratory, California Institute of Technology, under a contract with the National Aeronautics and Space Administration (80NM0018D0004). This research has made use of data obtained with \textit{NuSTAR}, a project led by Caltech, funded by NASA and managed by the NASA Jet Propulsion Laboratory (JPL), and has utilized the \textsc{nustardas} software package, jointly developed by the Space Science Data Centre (SSDC; Italy) and Caltech (USA). This research has also made use of data obtained with \textit{XMM–Newton}, an ESA science mission with instruments and contributions directly funded by ESA Member States. 
\section*{Data Availability}

All the data utilized in this article are publicly available through ESA's \textit{XMM-Newton} Science Archive (\url{https://www.cosmos.esa.int/web/xmm-newton/xsa}) and NASA’s HEASARC archive (\url{https://heasarc.gsfc.nasa.gov/}).



\bibliographystyle{mnras}
\bibliography{example} 








\bsp	
\label{lastpage}
\end{document}